\documentclass[twocolumn]{elsartUSA}

\usepackage{graphicx}
\usepackage{float}
\usepackage{amssymb}
\usepackage{color}
\usepackage[bbgreekl]{mathbbol}
\usepackage{bm}
\usepackage{soul}
\usepackage{multirow}
\usepackage{hhline}
\def \Ca{C_\alpha}

\journal{Journal of Molecular Biology}

\begin{document}

\begin{frontmatter}

%% Title, authors and addresses

%% use the tnoteref command within \title for footnotes;
%% use the tnotetext command for the associated footnote;
%% use the fnref command within \author or \address for footnotes;
%% use the fntext command for the associated footnote;
%% use the corref command within \author for corresponding author footnotes;
%% use the cortext command for the associated footnote;
%% use the ead command for the email address,
%% and the form \ead[url] for the home page:
%%
% \title{Title\tnoteref{label1}}
%% \tnotetext[label1]{}
%% \author{Name\corref{cor1}\fnref{label2}}
%% \ead{email address}
%% \ead[url]{home page}
%% \fntext[label2]{}
%% \cortext[cor1]{}
%% \address{Address\fnref{label3}}
%% \fntext[label3]{}

\title{Mechanical probes of SOD1 predict
  systematic trends in metal and dimer affinity of ALS-associated
  mutants \\}

%% use optional labels to link authors explicitly to addresses:
%% \author[label1,label2]{<author name>}
%% \address[label1]{<address>}
%% \address[label2]{<address>}

\author{Atanu Das and Steven S. Plotkin${}^{\ast}$}
% \author[1]{Atanu Das}
% \author[2]{Steven S. Plotki\corref{Corresponding author:}}
% \ead{steve@phas.ubc.ca}

%Corresponding author: steve@physics.ubc.ca

\address{Department of Physics and Astronomy, University of British
  Columbia, Vancouver, B.C. V6T 1Z1.  \hspace{1cm} ${}^{\ast}$\tt{Corresponding author:
    steve@physics.ubc.ca} }

\begin{abstract}
%% Text of abstract
\scriptsize{Mutations and oxidative modification in the protein Cu,Zn superoxide
dismutase (SOD1) have been implicated in the death of motor neurons in
amyotrophic lateral sclerosis (ALS), a presently incurable, invariably
fatal neurodegenerative disease. Here we employ steered, all-atom
molecular dynamics simulations in implicit solvent to investigate the significance
of either mutations or post-translational modifications (PTMs) to SOD1
on metal affinity, dimer stability, and mechanical malleability. The
work required to induce moderate structural deformations as a function of
sequence index constitutes a ``mechanical fingerprint'' measuring
structural rigidity in the native basin, from which we are able to
unambiguously distinguish wild-type (WT) SOD1 from PTM variants, and measure the
severity of a given PTM on structural integrity. The cumulative
distribution of work values provided a way to cleanly discriminate
between SOD1 variants. 
Disulfide reduction destabilizes dimer stability more than the
removal of either metal, but not moreso than the removal of both metals.  
Intriguingly, we
found that disulfide reduction mechanically stabilizes apo SOD1 monomer,
underscoring the differences between native basin mechanical
properties and equilibrium thermodynamic stabilities, and elucidating
the presence of internal stress in the apo state. All PTMs and
ALS-associated mutants studied showed an increased tendency to lose
either Cu or Zn, and to monomerize- processes known to be critical in
the progression of ALS. The valence of Cu strongly modulates its
binding free energy. As well, several mutants were more susceptible
to loss of metals and monomerization than the 
disulfide-reduced or apo forms of SOD1.
% Free energies for metal binding and
% dimer separation are obtained using the weighted histogram analysis
% method (WHAM), with distance constraints to prevent
% large-scale structural deformation, followed by re-equilibration after
% the respective process. When distance constraints are removed,
% the WHAM results agree with those obtained from direct application of
% the Jarzynski equality.
Distance constraints are required to calculate free energies for metal
binding and dimer separation, which are validated using thermodynamic
cycles. When distance constraints are removed, the results agree with
those obtained from direct application of the Jarzynski equality.}

\end{abstract}

\begin{keyword}
%% keywords here, in the form: keyword \sep keyword
superoxide dismutase \sep ALS \sep molecular dynamics simulations \sep
thermodynamic stability \sep protein misfolding
%% MSC codes here, in the form: \MSC code \sep code
%% or \MSC[2008] code \sep code (2000 is the default)

\end{keyword}

\end{frontmatter}

%%
%% Start line numbering here if you want
%%
%\linenumbers

%% main text
\section{\scriptsize{Introduction}}
\label{intro}

\scriptsize{Copper, zinc superoxide dismutase (SOD1) is  a homo-dimeric
antioxidant enzyme of 32kDa present in all eukaryotes.
Each monomer consists of an eight stranded greek key $\beta$
barrel of 153 amino acids~\cite{TainerJA82,BertiniI98,ValentineJS05} and binds one Cu and one
Zn ion, which significantly enhance native thermodynamic and
mechanical stability~\cite{ShawBF2007,DasA12pnas1}.

Two large, functionally important loops determine the structure and
activity of the enzyme. Loop VII or the electrostatic loop (residues
121-142) contains charged and polar residues that enhance enzymatic
activity by inducing an electrostatic funnel towards
the active site centered on the Cu ion~\cite{GetzoffED92}, which
catalyzes the conversion of superoxide (O${}_2^-$) to less toxic
species(either O${}_2$ or H${}_2$O${}_2$). Loop IV or the Zn-binding
loop (residues 49-83) coordinates Zn with histidines 63, 71, 80 and
Aspartic acid 83, which, along with a disulfide bond between Cysteines
57 and 146, enforce concomitant tertiary structure in the
protein~\cite{RobertsBR07}. Residues in the Zn binding loop form about
38\%  of the dimer surface contact area, so that disorder in the loop
due to Zn expulsion and/or disulfide reduction can facilitate
monomerization of the
homodimer~\cite{ArnesanoF04,HoughM2004,RakhitR04,HornbergA07,KayatekinC2010}.

The Cu on the other hand is coordinated by residues mainly in the
$\beta$ strands of the immunoglobulin-like core of the protein-
histidines 46, 48, 120, along His 63 which bridges the Cu and Zn ions-
so that protein structure only weakly couples to Cu binding~\cite{BanciL03,StrangeRW03},
supporting a primarily enzymatic role of Cu.

Amino acid missense mutations at more than 150 positions in SOD1 have been
found to cause amyotrophic lateral sclerosis (ALS), an invariably
fatal neurodegenerative disease characterized by loss of
the motor neurons in the brain, brainstem and spinal cord
(http://alsod.iop.kcl.ac.uk/). Such familial mutations constitute about
20\% of the cases of ALS known to display autosomal
dominance~\cite{RosenDR93}, conferring symptoms
through a
mechanism involving a toxic gain of function, in that SOD1 knockout
mice do not develop the neurodegeneration indicative of the familial
ALS (fALS)-like phenotype~\cite{ReaumeAG96}.
Gain of function symptoms have been variously attributed to generation of reactive oxygen
and nitrogen species, cytoskeletal disruption, caspase activation,
mitochondrial dysfunction, proteosome disruption, and microglial
activation~\cite{ClevelandDW01,HartP2006,RakhitR2006}.
SOD-mediated
fALS is one of the most prominent identified causes of the disease
despite constituting only approximately 2-5\% of all known cases.
The vast majority, about 90\%, of ALS cases have no known underlying
etiology and are termed sporadic (sALS). Macroscopically, sALS is clinically
indistinguishable from fALS~\cite{GaudetteM00}. Lewy body-like inclusions in sALS
have been found to be immunoreactive to SOD1-specific antibodies~\cite{ShibataN94},
and  misfolding-specific antibodies have also identified misfolded SOD1 protein
in sALS inclusions~\cite{BoscoDA10,ForsbergK10}.
At the molecular level however, protease resistant cores in aggregates as identified by
MALDI-TOF mass analysis were observed to differ between the WT sequence and the
fALS-associated mutants G37R, G85R, and G93A~\cite{FurukawaY10}, indicating that
mutations can modulate structural polymorphism in aggregate cores,
resulting in distinct physico-chemical aggregate properties such
as sequence-dependent solubility. The mutation-induced polymorphism of
fibril cores has been recapitulated in coarse-grained simulations
using discrete molecular dynamics with
structure-based, G\={o}-like potentials~\cite{DingF12}
Nevertheless, several studies have indicated oxidative modification of WT SOD1 can
be toxic in ALS, linking SOD1 misfolding to a common pathogenic mechanism in fALS
and sALS~\cite{EstevezAG99,RakhitR02,GruzmanA07,EzziSA07,KabashiE07}.

Studies involving misfolding-specific antibodies have shown that
misfolded SOD1 (either G85R or G127X) can induce misfolding in
natively-structured WT SOD1 by direct protein-protein
interactions~\cite{GradLI11}.  Holo, pseudo WT SOD1 (C6A/C111S) has
been observed to aggregate under physiologically relevant conditions,
with characteristics similar to aggregates in fALS
patients~\cite{HwangYM10}.  Further, cell to cell transfer mediated by
macropinocytosis has been observed in the fALS mutant
H46R~\cite{MunchC11}.  These studies point to prionogenic mechanisms
for the propagation of SOD1 misfolding and aggregation- a common theme
in protein-misfolding diseases.

SOD1 protein need not globally misfold or unfold in order to aggregate. 
Accumulated evidence for several proteins indicates that aggregation
may be initiated from locally (rather than globally) unfolded
states~\cite{ChitiF09}. These partially disordered states may be induced by external
agents or may become accessible via thermal fluctuations or rare
events. For the case of SOD1, a near native
aggregation precursor was found for an obligate
monomeric SOD1 variant (C6A/C111A/F50E/G51E), wherein protective cap
regions are locally unfolded around the native $\beta$-stranded
core~\cite{NordlundA06}. As well, the crystal structures of the fALS mutants
S134N and apo H46R have both been observed to adopt
fibrillar structures with intercalating loops between largely
native-like domains~\cite{ElamJS03nsb}, and  S134N
has been observed to form
oligomers in solution that are stabilized by locally unfolded elements from the
native structure, involving transient
interactions between electrostatic loops from different dimers~\cite{BanciL2005}.

The above studies have motivated a previous computational study, where
we had focused on the native and near-native mechanical properties of
WT SOD1, premature variants lacking post-translational modifications,
and both ALS-associated and rationally designed truncation
mutants~\cite{DasA12pnas1}.  We felt that this approach was
complimentary to experimental structural and thermodynamic assays.  In
that work, we examined relatively large changes in native structure as
a result of significant perturbing mechanical forces, however the
forces are not so large as to induce global unfolding. The mechanical
forces are applied across the whole protein surface to ascertain a
``mechanical fingerprint'' for a given SOD1 protein variant.

An all-atom, implicit solvent model was used, which in benchmark cases
compared favorably to an explicit solvent model, but less favorably
with a structure-based G\={o} model.
The (non-equilibrium) work values obtained by pulling a residue to 5\AA~
were found to strongly correlate ($r=0.96$) with the equilibrium free energy change for the
same process, as calculated using the weighted histogram analysis
method (WHAM). Thus the work profiles obtained accurately represented
the relative thermodynamic stability of various regions, or the
relative thermodynamic stability between mutant and WT for the same
region.

We found
that the mechanical profile of a given SOD1 variant could be best
represented through the cumulative distribution of mechanical work
values, where the work is obtained by pulling various residues in the
protein out to a given distance. The cumulative distribution of work
values can be thought of as a generalization of native ``rigidity'',
which accounts for the fact that the rigidity can vary place to place,
so that the collection of work values themselves obey a distribution
that differs between SOD1 variants.
For purposes of distinguishing cumulative work distributions of SOD1
variants, we saw that mechanical data collected for 48 residues was
sufficient to represent more comprehensive sets of data, in that the cumulative work
distribution seemed to have converged to within $\approx 1$ kJ/mol
after a sample size of about 40 residues.

Mechanical fingerprinting studies of SOD1 variants with Zn-binding and
electrostatic loops either truncated to Gly-Ala-Gly linkers~\cite{DanielssonJ11},
or extended by poly-Glycine insertions
revealed that, although the Zn-binding and electrostatic
loops in apo SOD1 destabilized the $\beta$-sandwich
core of the protein in the absence of metals, evolution has responded by strengthening
interactions in regions flanking the loops, to preserve the structural
integrity of the core domain and presumably prevent misfolding.
Nevertheless, mechanical fingerprinting studies of a series of C-terminally truncated
mutants, along with an analysis of equilibrium dynamic fluctuations
and solvent exposure while varying native constraints, together revealed
that the apo protein is internally frustrated, and that this
internal strain is an allosteric consequence of evolution towards high
metal-binding affinity; release of the stress in a truncation mutant
causes loss of metal binding function, even though all metal-binding
ligands are still present~\cite{DasA12pnas1}.
Thus, the evolutionary optimizaton of SOD1 function as a metal-binding
redox substrate competes
with apo state thermo-mechanical stability, consequently increasing the
susceptibility of misfolding in the apo state.

Here we examine mechanical properties of mutant and WT SOD1,
as well as the free energetic changes accompanying
post-translational modifications (PTMs) such as metal binding and
dimerization.   The affinity for
metals as well as dimer stability are determined both by weighted
histogram analysis methods as well as direct mechanical pulling assays
using the Jarzynski equality, and the methods are checked for
consistency. Of the 21 ALS-associated mutants investigated here, 
every one showed both reduced metal and dimer affinity. 
Some mutants had lower dimer affinity in the holo state than WT
protein in the apo, disulfide reduced state. PTMs
such as disulfide reduction or metal depletion also lower dimer stability
and metal affinity, e.g. Cu-depletion lowers the affinity for Zn and
vice-versa. Zn plays a larger role than Cu in determining the mechanical rigidity
of the native state. Presence of the native C57-C146 disulfide bond
induces stress in the apo state of the protein, which is relieved
either upon metal binding, or reduction of the bond. Thermodynamic
quantities as 
obtained from our simulations are compared with those obtained
experimentally in the Discussion section.}

\section{\scriptsize{Results}}

\subsection{\scriptsize{The Jarzynski/Crooks equality overestimates the free energy
  change, due to conformational distortion}}
\label{jarzynskiF}

\begin{figure}[H]
\includegraphics*[width=8cm]{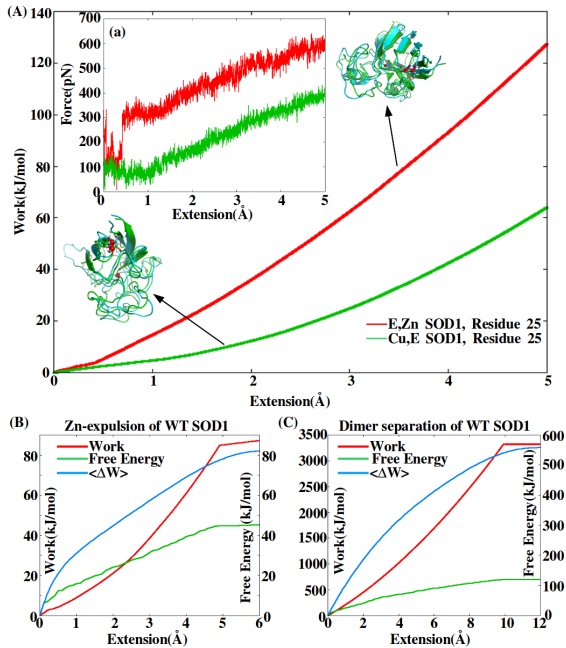}
\caption[]{
\tiny{(Panel A): Work performed as a function of distance pulled. Tethers are placed at
the C$_\alpha$ atom closest to the center of mass of the SOD1 monomer
(H46), and the C$_\alpha$ atom of residue S$25$ of either (E,Zn) (red) or (Cu,E)
(green) SOD1, in separate pulling assays. Inset (a)
Force-extension profile of  residue $25$ for (E,Zn) (red) and (Cu,E)
(green) SOD1. Ribbon representations of the protein are also shown; the
tethering residue is shown in licorice rendering (in red) and the
center C$_\alpha$ as a red sphere. The initial equilibrated (at
$0$~\AA~, green ribbon) and final (at $5$~\AA~, blue ribbon) structures are aligned to each other by
minimizing RMSD; this indicates that overall, the structural perturbation
induced by pulling one residue to $5$~\AA~ is modest, but significant
enough to be unlikely thermodynamically based on the magnitude of the work values.
(Panel B): Work and WHAM-derived free energy to move Zn a given distance from its
putative position in WT SOD1. Values have nearly converged after 5\AA.
Note that
the free energy is not always less than the work for a given
trajectory (red);
the thermodynamic inequality $\left< \Delta W\right> > \Delta F$ only
holds true after ensemble averaging, as seen for the blue curve
representing the ensemble-averaged work (over 20 trajectories)
to extend the metal to a given distance.
(Panel C): Work (red), average work (blue), and WHAM-derived free
energy (green) to separate the WT SOD1 dimer by a
given separation distance. Values have nearly converged after 10\AA. The free energy is substantially less than the work, much of which
goes into conformational distortion of the protein when harmonic
constraints on $C_\alpha$ atoms are not present.}
}
\label{figexample}
\end{figure}

\scriptsize{Figure~\ref{figexample}A plots the work performed as a function of
extension, for residue 25 of two SOD1 variants,  (E,Zn) (Cu-depleted)
SOD1 and (Cu,E) (Zn-depleted) SOD1. The protocol is described in the Methods
Section~\ref{sectsteeredmd}. 
Some variants have softer
mechanical profiles than others; for residue 25, (Cu,E) is softer than
(E,Zn). Figure~\ref{figexample}B shows the work to pull the Zn
ion to a given distance from its putative position in WT protein,
along with the free energy as obtained from 
the WHAM method, as described in Methods
section~\ref{whammethoddimermetal}. 
Note that the free energy is not always less than the
work for this trajectory- it is only when ensemble-averaged that the
thermodynamic inequality
$F \leq \left< W \right>$ holds. Both work and free energy begin to
converge to their asymptotic values at
distances beyond about 5 \AA.  Figure~\ref{figexample}C shows the
work to monomerize the WT dimer by pulling it apart, along with 
the free energy change as obtained from the WHAM method. Note that
the work performed is substantially larger than the free energy
change. Both work and free energy begin to
converge to their respective asymptotic values after about
10 \AA. The methods used for dimer separation are described in
Sections~\ref{sectsteeredmd} and~\ref{whammethoddimermetal}. 

To calculate the free energy of local protein unfolding, metal expulsion, 
and dimer separation, we have calculated the work to either pull a particular $\Ca$ to
$5$\AA, pull a metal ion until the force drops to zero, or pull
monomers in the dimer apart until the force drops to zero.
The various assays are repeated multiple times, and the 
corresponding free energy changes are obtained from the work 
values by applying the
Jarzynski equality~\cite{EvansDJ94,JarzynskiC97,CrooksG98}. Finite
sample-size corrections are accounted for~\cite{GoreJ03}, as described in Methods
section~\ref{freejarzynski}. 

The free energy change is calculated from 
repeated pulling assays measuring the non-equilibrium work
value to pull a residue to 5\AA, through the Jarzynski equality:
$\mbox{e}^{-\Delta F/kT} = \left< \mbox{e}^{-W/kT}\right>$. Here $\Delta F$
is the free energy change, $W$ is the work value, $kT$ is
Boltzmann's constant times the temperature in Kelvin, and the
brackets $\left< \cdots \right>$ denote the ensemble average over
identical pulling assays.} 

\begin{figure}[H]
\includegraphics*[width=8cm]{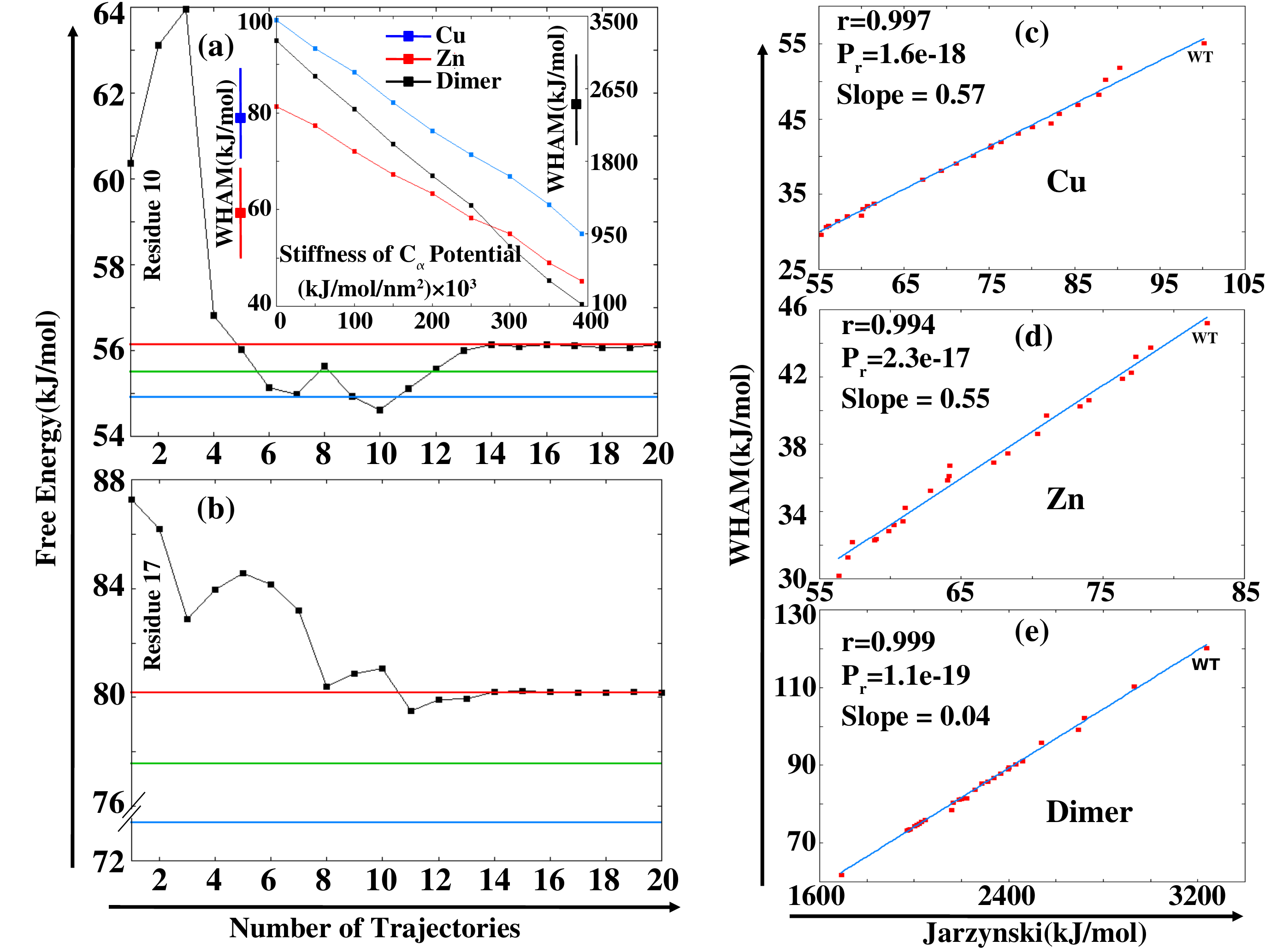}
\caption[]{
\tiny{Comparison of free energy values calculated
from Jarzynski's equality, and from the weighted histogram analysis
method (WHAM).
(Panel a) Free energy as calculated using Jarzynski's equality, as a
function of the number of repeated pulling assays or trajectories, for
residue $10$ to be pulled to $5$\AA~ from its equilibrium position in
a direction radially outward from the center of mass. (Panel b) The
same plot as panel (a) for reside $17$. In panels (a) and (b), red
horizontal line indicates the value of free energy calculated
from the Jarzynski
Equation~(\ref{eqjn}). The
green line indicates the free energy obtained by accounting for extra corrections arising
from finite sample sizes~\cite{GoreJ03} described in Method section~\ref{freejarzynski}.
The blue line indicates
the free energy value calculated by WHAM, wherein equilibration with umbrella
sampling constraints allows conformational distortion to relax, see Methods section~\ref{whammethodmech}.
(Panel c) Scatter plot of the free energy values for Cu removal,
as calculated by the
Jarzynski equation~(\ref{eqjn}) and WHAM with $\Ca$ constraints, for the
mutants give in Table~\ref{tabwham}. WT protein in the upper right of the
plot is indicated, and has the largest binding free energy. The free energy values
strongly correlate between the two methods. (Panel d) Same as Panel
(c) for the removal of Zn.
(Panel e) Same as Panel (d), for the free energy of dimer separation.
Note that although the free energies obtained by two methods strongly
correlate, the slope of the best fit line is much less than unity,
indicating the presence of substantial additional free energy changes due to
conformational distortion using the Jarzynski method.
(Inset to panel a) WHAM-derived free energy changes for metal
expulsion and dimer separation for WT holo SOD1, as a function of the stiffness of the
harmonic constraints on the $\Ca$ atoms. The rightmost points on each curve
correspond to the WHAM free energies in panels (c-e), i.e. at the values of harmonic
constraint used to generate the WHAM free energies for those
plots. The WHAM free energies at zero stiffness are consistent with
the Jarzynski values of the free energy in panels (c-e).
}}
\label{figJarWhamCorr}
\end{figure}

\scriptsize{The results are shown in Figure~\ref{figJarWhamCorr}: Panels (a)
and (b) show plots of the quantity  $-kT \ln \left<
  \mbox{e}^{-W/kT}\right>_N$, where the above ensemble average is
evaluated over $N$ trajectories (identical runs), as a function of the
number of trajectories $N$. For sufficiently large $N$, the average should converge
to the ensemble average. The plots in Figure~\ref{figJarWhamCorr}
indicate that about $14$ trajectories is sufficiently large for convergence, for the
pulling rates and extensions we considered in our study (red
horizontal lines in Panels (a) and (b) of
Figure~\ref{figJarWhamCorr}.} 

\scriptsize{Finite sample size effects reduce the estimate for the free energy
change, however the mean dissipated work in our pulling assays is modest, only about a kJ/mol,
indicating} 

\renewcommand{\arraystretch}{1}
\begin{table}[H]
\scriptsize
\caption{Proteins considered for mechanical scan, metal-affinity, and dimer-stability estimation}
\begin{tabular}{c|c}
\hline
\multicolumn{2}{c}{Mechanical scan${}^a$/Metal expulsion/Monomerization} \\
\hline
SOD1 variant & PDB ID used for \\
& structure generation \\
\hline
WT holo & 1HL5, 2C9V\\
(Cu,E) & 2R27${}^{b,c}$ \\
(Cu-shift,E) & 2R27${}^{b,c}$ \\
(E,Zn) & 1HL4\\
apo(SH)${}^d$ & 1HL5, 1RK7${}^c$ \\
apo${}^d$ & 1RK7${}^c$ \\
holo(SH) & 2AF2${}^c$ \\
\hline
G127X & 1HL5${}^e$ \\
A4V & 1UXM \\
D124V & 3H2P \\
D125H & 1P1V \\
D76Y & 1HL5${}^e$ \\
D90A & 1HL5${}^e$ \\
G37R & 1AZV \\
G41D & 1HL5${}^e$ \\
G41S & 1HL5${}^e$ \\
G85R & 2ZKW \\
G93A & 3GZO \\
G93C & 1HL5${}^e$ \\
H43R & 1PTZ \\
H46R & 2NNX${}^c$ \\
H46R/H48Q & 2NNX \\
H80R & 3QQD \\
I113T & 1UXL \\
L144F & 1HL5${}^e$ \\
L38V & 2WZ5 \\
S134N & 1OZU \\
T54R & 3ECW \\
W32S & 1HL5${}^e$ \\
\hline
\label{tabwham}
\end{tabular}
\\
${}^a$Mechanical scans were performed for SOD1 variants
above the horizontal line located between holo(SH) SOD1 and G127X. \\
${}^b$Missing regions in the PDB structure were remodeled as
described in section~\ref{remodeling}.\\
${}^c$Structures reported have mutations from the WT sequence;
these were ``back-mutated'' to construct structures for the WT
sequence, as described in Methods
sections~\ref{remodeling},\ref{modelingnopdb}. \\
${}^d$Metal expulsion is not relevant for these variants.  \\
${}^e$These mutants have no PDB structure reported, so are
constructed by mutating the appropriate residues in PDB 1HL5. \\
\end{table}

\scriptsize{that pulling is near equilibrium, and thus finite
sample-size corrections are not large: about 1-3 kJ/mol (green
horizontal lines in Panels (a) and (b) give the refined estimate for
the free energy change accounting for finite sample-size
corrections). However, the free energy changes obtained from this method are still larger
than those of the WHAM method (blue horizontal lines in Panels (a) and (b) of
Figure~\ref{figJarWhamCorr}). The discrepancy is about $0.6$ kJ/mol
when residue 10 is pulled to 5 \AA, and about $5$ kJ/mol 
when residue 17 is pulled to 5 \AA. 
The reason is that pulling on a particular residue at a finite rate distorts the rest
of the protein, and there is a free energy change corresponding to
this distortion that contributes to the total. The distortion may be avoided by applying constraints,
but it is not obvious {\it a priori} how or where to apply such
constraints when pulling residues away from the protein. 
The WHAM method minimizes the free energy change due to
distortion by equilibrating at each distance ``window'' in the
corresponding umbrella potential. 
Convergence was tested by varying the number of windows between 25 and
40, and varying the equilibration time within each window between 10ns
and 25ns, for which the results did not noticeably change. 

The free energy changes for metal expulsion and dimer separation may
also be calculated from both the Jarzynski equality and WHAM
method. 
For these assays, distortion of the protein may be minimized by
applying harmonic constraints to the $\Ca$ atoms of the protein. 
The harmonic constraints are applied in the WHAM analysis, and
the results compared with those of the (unconstrained) Jarzynski
analysis. 
Side chains and other backbone atoms are allowed to move. 

Perhaps the main concern is whether applying such constraints would
impede the removal of the metal ion. The most direct,
solvent-accessible pathway for metal expulsion was taken, as described
in Methods section~\ref{secttethering}. 
To confirm that metal
expulsion was not impeded by $\Ca$ atom constraints, the stiffness of
the $\Ca$ potential was varied, from zero (unconstrained) to
$392\times 10^3$ kJ/mol/nm${}^2$, the stiffness of a single carbon-carbon bond,
as used in the WHAM method.  
If constraints impeded the process of metal
expulsion, a minimum would be seen as a function of stiffness
of the constraining potential. If constraints did not impede the
process, the free energy vs. the stiffness of the constraining
potential would be monotonically decreasing. 

The WHAM free energy of metal expulsion and dimer separation 
for the WT protein as a
function of the stiffness of the constraining potential is plotted in
the inset to Panel (a) in Figure~\ref{figJarWhamCorr}. 
The free energy is indeed monotonically decreasing, indicating that
the constraints do not impede the process of metal expulsion or dimer
separation. The free energy of deformation may thus safely be minimized by
applying such constraints, to obtain the direct free energy cost of
metal expulsion or dimer separation. 

Figure~\ref{figJarWhamCorr}, Panels
(c-e), plots the free energy changes for Cu expulsion, Zn expulsion, and dimer separation for the
mutants of SOD1 given in Table~\ref{tabwham}. WHAM values are plotted on
the ordinate; Jarzynski values, as given by
equation~(\ref{eqjn}) for $25$ replicas (i.e. without any finite
size-corrections), are plotted on the abscissa.  
The unconstrained Jarzynski method overestimates the free energy
changes, for the reasons above. The slopes of the best fit lines are all less
than unity, and for dimer separation the slope is particularly
small. However the correlations between the two separate methods are
excellent, indicating that the deformation free energy is not a large
and random component of the total free energy, but likely correlates
with the direct metal expulsion or monomerization free
energy.

For both Zn and Cu binding, the ratio of the binding free energies, Jarzynski to
WHAM, is about 1.8.
Relative rank ordering of the free energies between the WT
sequence and mutants may be predicted either by the WHAM or the Jarzynski
method. We found, however, that the WHAM method was the most rapid and
reliable way of determining free energy changes in the system.

The right-most data point in Panels (c-e) refers to the WT protein (labelled),
i.e. the WT protein has the largest metal and dimer affinity. This is
described in further detail in
Sections~\ref{metalaffinity},~\ref{sectmonomerize} below.  
Comparing the WHAM and Jarzynski values for say Cu expulsion in Panel
(c) of Figure~\ref{figJarWhamCorr}, the WHAM value of
$55$ kJ/mol is taken from the value in the inset to panel (a) at
$\Ca$ constraint of $392\times 10^3$ kJ/mol/nm${}^2$, 
while the Jarzynski value of $100$ kJ/mol is consistent with the
unconstrained WHAM value of $99$ kJ/mol in the inset of Panel (a) 
(accounting for finite-size corrections to the Jarzynski estimate
reduces the value by only about $0.3$ kJ/mol). 
Thus, the WHAM
values approach the 
Jarzynski values as the $\Ca$ constraints are removed, and provide a
satisfying consistency check.}

\subsection{\scriptsize{ALS-associated SOD1 mutants show decreased affinity
  for both Cu and Zn}}
\label{metalaffinity}

\scriptsize{Copper and Zinc ions were pulled out of their putative binding sites
for the list of proteins given in Table~\ref{tabwham}. For mutant
proteins with no PDB structure, 
the initial structure was generated by
mutating the appropriate residue(s) in the WT structure (PDB 1HL5)
and equilibrating, according to
Methods section~\ref{modelingnopdb}. 
Some mutants of SOD1 are devoid of Cu in the PDB structure- these are H46R,
H46R/H48Q, S134N, D124V, H80R, and D125H, and some are fully
metal depleted: G127X and
T54R~\cite{BanciL09,JonssonPA04,GradLI11}. 
For these proteins, metals were incorporated into the proteins by
superposing them onto holo SOD1. The metals were found to be
metastable in their binding sites, i.e. no forces were required to constrain the
metals to their putative bound positions for the duration of the
simulation.

The tethering
residues are chosen according to Methods section~\ref{secttethering},
to select an easy pathway for metal expulsion;
the other tethering point is the metal itself.  From these pulling
assays, 25 configurations were taken at
separations between $0$ and $5$ \AA, 
and used as initial states in umbrella sampling, to obtain the free
energy to extract the metal using WHAM. 

The free energy values of metal expulsion thus obtained are plotted in
Figure~\ref{figCuZnFreedimer} (a), and also given in Table~S1. %\ref{tabfree}.
The values in Figure~\ref{figCuZnFreedimer} (a) are sorted by decreasing Cu
affinity, and the values in Table~S1 are sorted by decreasing
Cu affinity separately amongst mutants and WT PTM variants. The WT
structure had the both the highest Cu and highest
Zn affinity, with values of $53$ kJ/mol and $40$ kJ/mol
respectively.

The free energy obtained by the WHAM method accounts for solvation
free energy (at the accuracy of the implicit solvent model), but not
fully for the chemical potential of the unbound metal, which is a
concentration-dependent quantity. The chemical potential must be taken
into account in determining the probability a metal is
bound. Nevertheless, it is clear that every mutation, whether the
corresponding amino acid was near the binding site or not, decreased
the affinity of the protein for the metals. A non-ALS associated
mutant, W32S, also reduced the affinity of the metals. The reduction
in binding free energy from WT SOD1 is up to
about $26$ kJ/mol for Cu (T54R) and $15$ 
kJ/mol for Zn (G127X). These
reductions are due to the induced
conformational strain at the binding site that arises from the resulting stresses of mismatching
interactions after a mutation. 

\begin{figure}[H]
\includegraphics*[width=6cm]{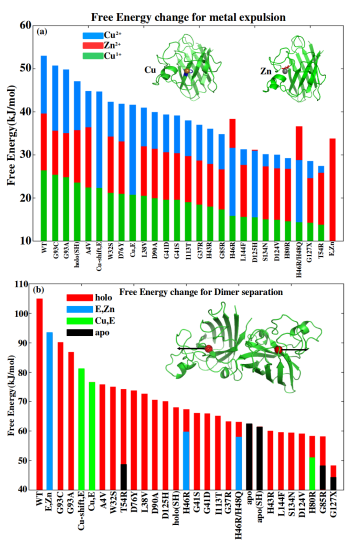}
\caption[]{
\tiny{(Panel a)
Free energy of copper (Cu) and zinc (Zn) binding/expulsion for WT SOD1 and
mutants, sorted by decreasing Cu affinity. Binding free energies are
obtained using WHAM as described in Methods section~\ref{whammethoddimermetal}.
Blue bars indicate Cu binding free energy, red bars indicate Zn
binding free energy, and green bars indicate monovalent Cu${}^{1+}$
binding free energy. The heights of all bars start from zero.
This binding energy accounts for solvation free
energy at the level of the implicit solvent model. Inset images are
intended to depict the direction of the pulling pathway, which was
chosen as the direction with the most immediate solvent exposure.
Metals are shown as an orange sphere for Cu and grey sphere for Zn, and the tethering
residue used in the pulling simulations is color coded directly behind it. For Cu
expulsion this was residue $45$ (blue cartoon), and for Zn expulsion
this was residue $83$ (red cartoon).
(Panel b)
Free energy to monomerize a homodimer of various SOD1 species. Both
ALS-associated mutants and variants lacking post-translational
modification (metal depleted or disulfide reduced) are
considered. Free energies are rank ordered strongest to weakest:
holo SOD1 has the strongest binding free energy, and G127X has the
weakest. All mutants and variants showed weaker dimer stability than
holo WT SOD1. The free energy to separate the dimer into monomers is obtained using
WHAM. The inset shows a ribbon representation of the monomers
constituting the dimer. The tethering residues, $\Ca(46)$ on each
monomer, are shown as red spheres, and the direction of pulling used
to generate initial conditions in WHAM is indicated by arrows.
All mutants are taken in the holo form, with Cu and Zn in their
putative binding positions. As well, H46R and H46R/H48Q do not bind
Cu, H80R does not bind Zn, and G127X does not bind either metal; in
these cases we also calculated the binding free energy for the
appropriate metal depleted forms. From Panel a, the Cu affinity for
the T54R variant was the lowest of all mutants, and the Zn affinity
was the 2nd lowest after G127X- lower than H80R which does not bind
Zn. For these reasons, we also calculated the free energy to monomerize apo
T54R dimer.
}}
\label{figCuZnFreedimer}
\end{figure}

\begin{figure}[H]
\includegraphics*[width=8cm]{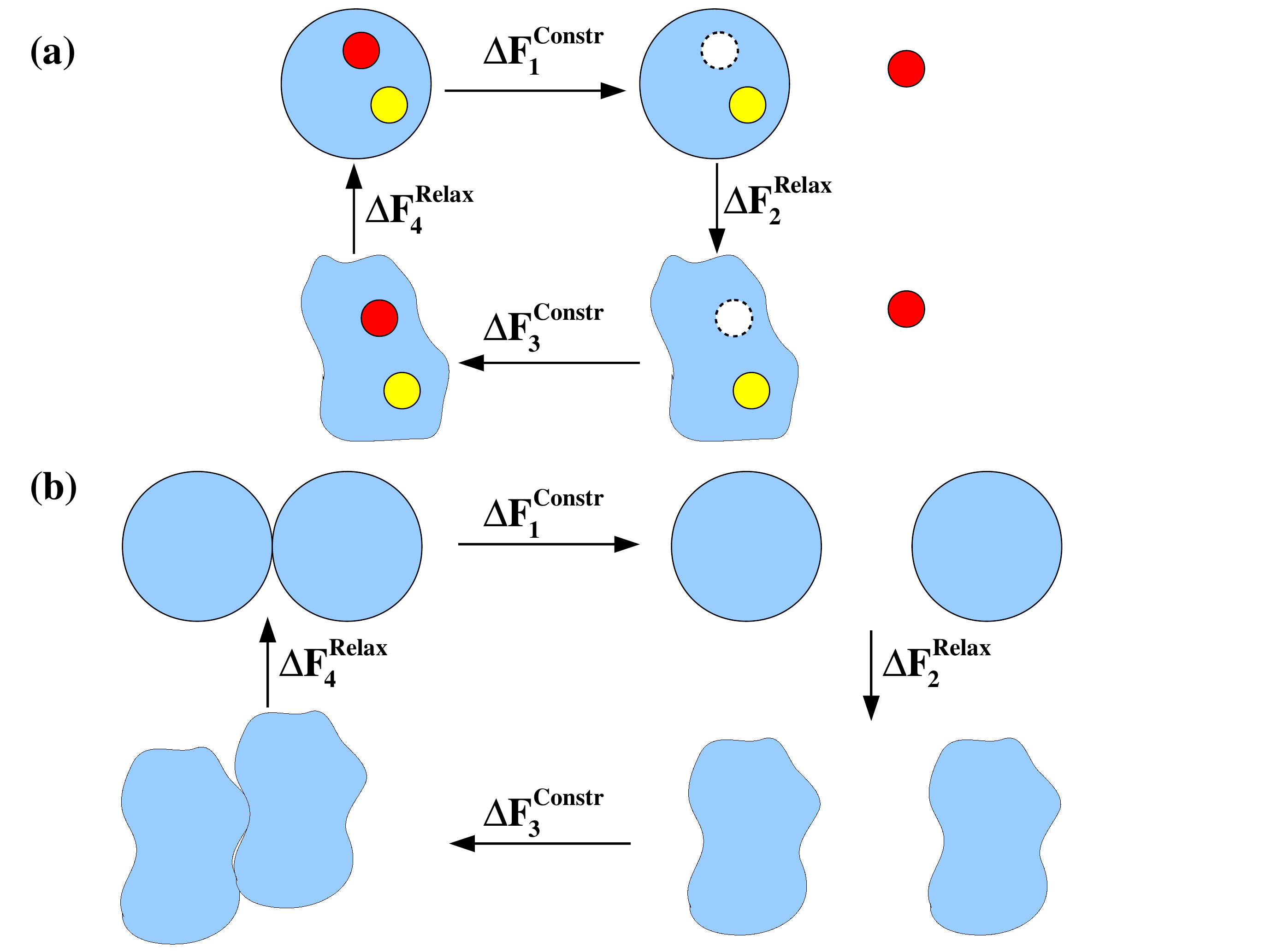}
\caption[]{
\tiny{Schematic representations of the thermodynamic cycles associated
with metal extraction and dimer separation of SOD1 variants. Panel (a):
Free energy changes associated with different steps of the
metal-expulsion/metal-insertion thermodynamic cycle. Cu and Zn ions
are shown as yellow and red spheres. $\Delta F_1^{constr}$ is the free energy change for metal
expulsion from the protein with the $\Ca$ constraints present (see text), $\Delta F_2^{relax}$
is the free energy change associated with the relaxation process where the
constraints were gradually removed after metal-expulsion, $\Delta F_3^{constr}$ is the
free energy change of inserting the same metal back into the half-metallated
protein, with $\Ca$ constraints present to preserve its initial backbone
conformation, and $\Delta F_4^{relax}$ is the relaxation
free energy after metal insertion.
The sum of the free energy changes around the thermodynamic cycle is
zero within the errors of the simulation, validating the calculation
of expulsion free energies.
Panel (b): Same as panel
(a), only here the free energy changes refer to monomerization or dimerization
compared to metal-expulsion or metal-insertion in panel (a).
}}
\label{figthermo}
\end{figure}

The free energy values of Cu and Zn expulsion are highly
correlated, i.e. if the affinity of Cu was reduced, so was that of
Zn. Three mutants stand out as exceptions: H46R, H46R/H48Q, and
D125H. Residues 46 and 48 coordinate the Cu ion, so the Cu affinity is
anomalously low in the corresponding mutants. 
The correlation coefficient between Cu${}^{2+}$ and Zn${}^{2+}$ binding free energies
without these 2 mutants was $0.929$, with significance $P_r =
4.4\mbox{e-}10$. The correlation coefficient between monovalent
Cu${}^{1+}$ and Zn${}^{2+}$ binding free energies 
without these 2 mutants was $r = 0.930$, with significance
$P_r = 3.7\mbox{e-}10$. 
The mean occupation numbers of Cu and Zn for several SOD1 mutants have
been obtained experimentally by Hayward {\it et
  al}~\cite{HaywardL2002}. Re-analyzing this data to inspect the
correlation between Cu and Zn content gives a correlation of $r=0.87$
($p=5\mbox{e-}6$), $r=0.84$ ($p=1\mbox{e-}4$) if mutants H46R, H48Q, and
D125H are removed.
The strong effect of D125H on Cu affinity is
more subtle and involves propagation of strain through the native
protein; further description of the effect is given in the Discussion
section.

SOD1 catalyzes the disproportionation of superoxide anion (O$_2^-$) in
a two-step reaction, wherein the Cu cation cycles between divalent and
monovalent states. In the monovalent state, the binding affinity for Cu${}^+$ is
reduced by slightly more than a factor of $2$, resulting in more
likely release of the cation, particularly for mutants. In fact, affinities for monovalent
Cu${}^+$ were always less than those of Zn${}^{2+}$, indicating ready
release of monovalent Cu upon structural perturbation or rare
thermodynamic fluctuation.

We have further checked that metal depletion and monomerization
each satisfy a thermodynamic cycle, such that upon metal re-insertion or 
re-dimerization, along with re-equilibration after the respective
process, the net free energy change over the cycle is 
zero within the errors of the simulations.
The thermodynamic cycles
are shown schematically in Figure~\ref{figthermo}, and
the free energy changes for the respective processes
are tabulated in Table~S2 of the Supplementary Content. 
On average $\sum_{cycle} \Delta F_i \approx 
0.15$ kJ/mol for metal expulsion/insertion, and 
$\sum_{cycle} \Delta F_i \approx 0.2$ kJ/mol for
monomerization/dimerization, for the cases that we benchmarked.
}

\subsection{\scriptsize{ALS-associated SOD1 mutants show increased tendency to
  monomerize}}
\label{sectmonomerize}

\scriptsize{Dissociation of the native SOD1 dimer is a prerequisite for its
aggregation~\cite{RakhitR04,KhareS2004,RumfeldtJAO06}. 
The pathway to aggregation generally proceeds by
non-native dimer formation and subsequent oligomer formation. We
investigated how both mutational and post-translational modification
modified dimer stability, for the ALS-associated mutants and PTM 
variants of SOD1 given in
Table~\ref{tabwham}. 

Using the WHAM methodology
described in Methods section~\ref{whammethoddimermetal}, we found the free
energy to separate the homodimer into a pair of monomers. A plot of
the dimer binding free energies, rank
ordered from strongest to weakest, is shown in
Figure~\ref{figCuZnFreedimer} (b).
The absence of any post-translational modification (e.g. absence of a metal,
reduction of the disulfide bond) lowered the free energetic
stability of the dimer. All ALS-associated holo mutants had 
reduced dimer stability, and several that we investigated had less dimer stability than that of
apo(SH) WT, which is known to monomerize~\cite{HornbergA07}. 
On the other hand, apo(SH) WT SOD1 had lower dimer stability than
  15/22 holo ALS-associated mutants. 
The non-ALS-associated mutant W32S showed reduced dimer stability. 
G127X, an obligate monomer in {\it in vitro} studies~\cite{GradLI11},
had the lowest dimer stability: a significant fraction of the 
residues participating in the dimer interface is removed by the
terminal sequence mutation. 

The trend in dimer stability did not correlate with the proximity of a
mutated residue to the dimer binding interface, either by measuring the
distance to the closest residue in the binding interface ($r=0.15$,$P=0.51$), 
or the mean
distance to residues putatively involved in the dimer interface:
residues 4, 50-53, 114, 148, 150-153 ($r=0.096$,$P=0.67$). These
results support a dimer destabilizing mechanism by mutations 
that involves the long-range propagation
of stress through interaction networks in the protein, 
an interpretation that is consistent with previous
experimental~\cite{VassallKA06} and simulation studies~\cite{KhareSD06pnas}.}

\begin{figure}[H]
\includegraphics*[width=8cm]{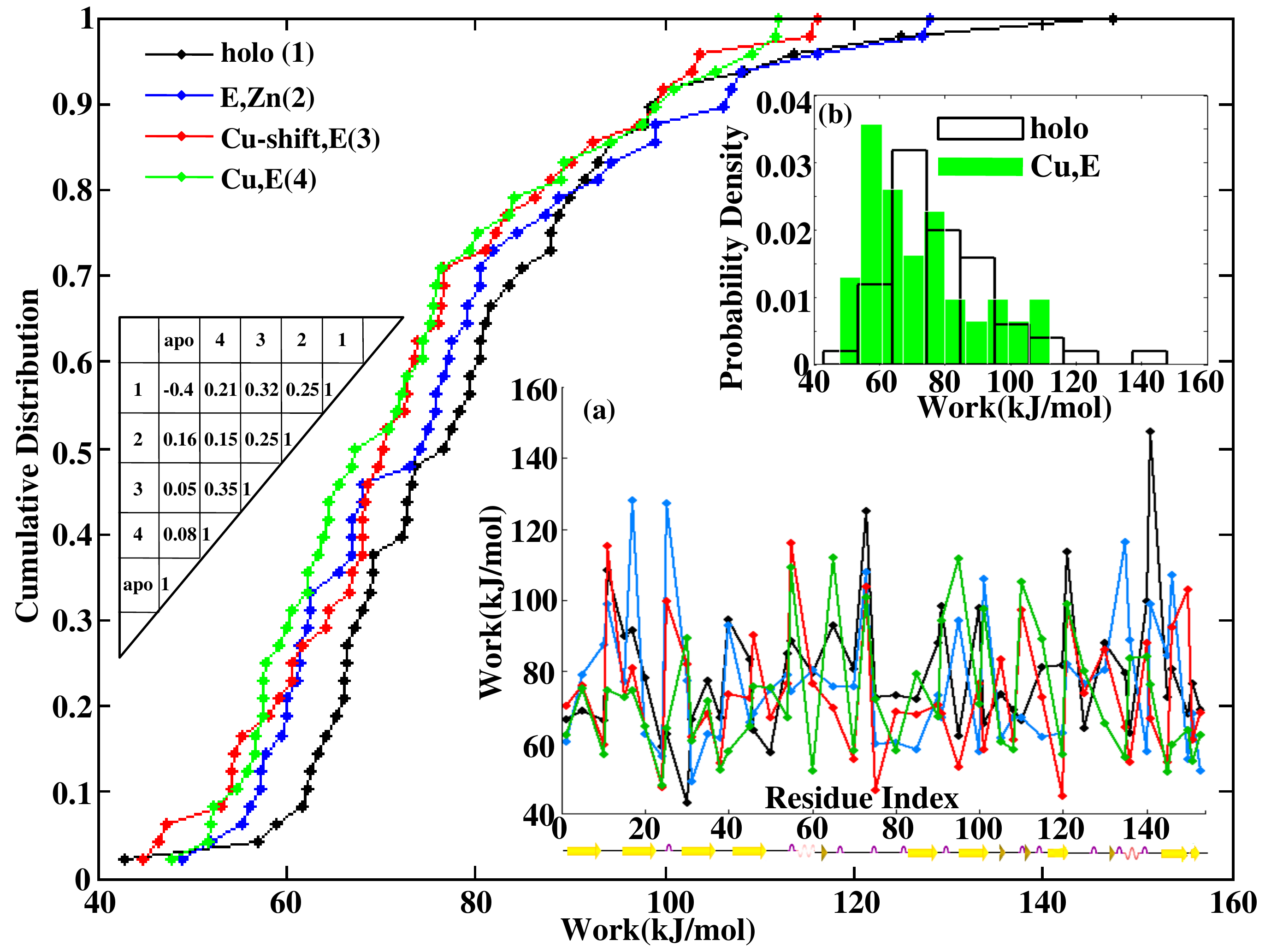}
\caption[]{
\tiny{The effect of metal depletion on the mechanical stability of
SOD1. (Inset panel a) Mechanical profiles of holo SOD1 (black), (E,Zn)
SOD1 (blue), (Cu,E) SOD1 (green), and (Cu,E) SOD1 with the Cu ion shifted
from its putative position to that of the Zn (red), which is the location of
the metal observed in the crystal structure of PDB
2R27~\cite{RobertsBR07}. Profiles are
color-coded according to the legend in the upper left. The effect of
metal depletion modulates the entire mechanical
stability profile.
Work values for individual residues are given in Table S5 of the
Supplementary Content.
(Inset b) Probability distributions of mechanical stability for holo
and (Cu,E) variants. This representation shows that the
distributions are indeed different, however there is significant
overlap, which along with the necessity of binning the data to
construct the histogram, makes the differences difficult to
quantify. (Main Panel) Cumulative distributions of the above SOD1
variants. The cumulative distribution requires no binning, and simply
rank orders and accumulates the work values. Using this
representation, discrepancies between work profiles most clearly
emerge. The mechanical stability of the SOD1 variants can be rank
ordered, strongest to weakest, as holo, (E,Zn),
(Cu-shift,E), and (Cu,E).
Also shown in the main panel is a correlation matrix resulting from a
pairwise comparison of the work values for all pairings of SOD1
variants. The  profiles do not correlate, again supporting the notion
that metal depletion modulates the entire mechanical stability
profile. A similar plot also
containing the cumulative distribution for apo SOD1 is given in the
Supplementary Content (Figure S1).
}}
\label{figEffectMetal}
\end{figure}

\subsection{\scriptsize{Metal depletion weakens the mechanical stability of WT
  SOD1-  Zinc moreso than Copper}}
\label{sectmetaldep}

\scriptsize{To investigate the effect of the presence or absence of Cu and Zn ions
on the mechanical stability of WT SOD1, we examined the mechanical
profiles of the metal present (holo), Cu-depleted (E,Zn), Zn-depleted (Cu,E), Zn-depleted with Cu
shifted to the Zn position (Cu-shift,E), 
and metal-depleted (apo) forms of SOD1. 
Mechanical profiles were obtained from pulling
simulations as described in Methods section~\ref{sectsteeredmd}.
Inset (a) of Figure~\ref{figEffectMetal} shows the
mechanical profiles of all the above variants except for apo
SOD1 (an analogous figure including the data for apo SOD1 is given in
Figure~S1 in the Supplementary Content).

The work values in the mechanical profile are significant compared to
thermal energies, ranging from about $16 k_{\tiny{B}}T$ to $56
k_{\tiny{B}}T$. Though the structural perturbations are fairly modest
in comparison to global unfolding, the energetics of the pulling
process is sufficiently non-perturbative that the work values across the
set of $48$ residues used here do not correlate with the RMSD values
for the respective residues (see Table~S3 in the Supplementary
Content). 

The mechanical profiles are significantly different for
all the variants- a table providing the cross-correlation
coefficients is given to the left of Figure~\ref{figEffectMetal}, 
and also in Table~S4 in the Supplementary Content, which
shows that no mechanical profiles is well-correlated with any other.
It is worth noting however that the correlation coefficients, though
weak, are statistically significant between some variants: The p-value
between (Cu,E) and (Cu-shift,E) is $0.015$, and the p-value between
holo and (Cu-shift,E) is $0.028$. 
The effect of demetallation is difficult to discern from the
mechanical profiles in inset (a). Here the cumulative distribution proves to be
valuable in elucidating the mechanical discrepancies between SOD1
variants, and is shown in the main panel of
Figure~\ref{figEffectMetal}. The probability distribution of work
values can differentiate work profiles for different SOD1 species
(inset (b) of Figure~\ref{figEffectMetal}), but unlike the cumulative
distribution it requires binning, and does not discriminate variants as clearly as the
cumulative distribution. 

According to the cumulative distributions in
Figure~\ref{figEffectMetal}, metal depletion mechanically destabilizes the
protein. That is, after depletion of either Cu or Zn, the protein is more
mechanically susceptible to perturbing forces. 
The distribution of apo SOD1
is broadened compared to (Cu,E) SOD1, but not overall
weakened (Figure S1, Supplementary Content). 
The relative importance
of Cu and Zn in determining the mechanical rigidity of the protein can
also be ascertained. Zn depletion results in more destabilization than
does Cu depletion, over the whole range of work values observed (green
and blue curves in Figure~\ref{figEffectMetal}).
That is, at any given perturbing work value, more residues in the (Cu,E)
protein would be substantially disordered than in the (E,Zn) protein. 
Equivalently, for say the weakest 1/3 (or any fraction) of the
residues, a smaller value of work is required to substantially
disorder those residues in the (Cu,E) variant than in the (E,Zn)
variant. Note that the residues in the weakest 1/3 may be different in
the two variants. 

The loss of Cu and/or Zn destabilizes both the Zn-binding loop and the
electrostatic loop, Zn moreso than Cu. Zn has close proximity to some of the residues
in the electrostatic loop, and is also involved in helix dipole
capping interactions with helix 133-138~\cite{GalaleldeenA2009}. 
From the work values in the mechanical profiles (See Table~S5 in the
Supplementary Content), the mean work for residues in the mechanical
scan that were in the Zn-binding loop (residues 50, 54, 55, 60, 65,
70, 73, 75, 80) was $83.9$ kJ/mol for holo, 
$78.2$ kJ/mol for (E,Zn), 
$76.4$ kJ/mol for (Cu,E), 
and $71.5$ kJ/mol for apo.
The mean work values for the corresponding
residues in the electrostatic loop (residues 121, 125, 130, 135, 136,
140, 141) was $93.5$ kJ/mol for holo, 
$85.6$ kJ/mol for (E,Zn), 
$77.7$ kJ/mol for (Cu,E) 
and $72.9$ kJ/mol for apo. 
The changes in work values with respect to holo SOD1 are largest
  for the zinc binding and electrostatic loops (Figure S7), indicating
  that the mechanical stability of these regions is preferentially
  dependent upon metal content. This finding is supported by dynamical
  fluctuation data, which show preferential increase in RMSF values
  for the zinc binding and electrostatic loops (Figure S7 and Table~\ref{tabvalandpredict}).

The changes in the mechanical stability profiles as a function of
sequence index, when metals are removed, are subtle to predict and
would likely involve a detailed quantification of the network of interactions
throughout the protein. The modulus of the changes in work values do
not correlate with simple parameters such as distance to the either metal
(all correlation coefficients had $r < 0.26$). 

In the crystal structure of (Cu,E) SOD1, Cu resides near the putative Zn
position. We tested whether this shift in position lowered the free
energy by pulling the Cu from its putative position in the crystal
structure to the Zn binding position, and applying umbrella sampling
with the WHAM method as described in Methods
section~\ref{whammethodshift}, to obtain the free energy change for
such a shift. Indeed these simulations gave
a reduction in free energy of $-5.7$ kJ/mol, indicating that the Zn binding position is
more favorable for the Cu ion, when only Cu is
bound.  However the
free energy barrier between the Cu and Zn binding positions is
sufficiently large, about $8$ kJ/mol, that we did not see the Cu change
binding positions during the course of our simulations (covering 0.2
$\mu$s) when Zn was absent. 
We tested the hypothesis that the shift of Cu also mechanically stabilizes the
protein, by calculating the mechanical
profile and cumulative distribution of (Cu-shift,E) % Zn-dep/Cu-shifted
SOD1. These results
are shown in Figure~\ref{figEffectMetal}, and the values are given in
Table~S5 of the Supplementary Content.
Shifting of the Cu ion to the Zn position
significantly increases in the mechanical stability of the
protein: The probability to obtain, by chance, a distribution as stable or more
stable than the (Cu-shift,E) 
distribution from the (Cu,E)  
distribution is $\approx 1.4$e-7 (see Methods
Section~\ref{statanal} and Figure S2 in the Supplementary Content). 
As mentioned above, the work profiles of (Cu,E) 
and (Cu-shift,E) 
SOD1 are only weakly correlated: 
$r=0.35$, $p=0.015$. 
(correlation table in Figure~\ref{figEffectMetal}). Shift of the Cu to the Zn position
globally changes the mechanical malleability of the protein.

The work profiles of both (Cu,E) 
SOD1 and (Cu-shift,E) 
SOD1 show lower stability in the metal binding residues H$80$ and
H$120$, compared to holo 
SOD1. The work values for residue $120$, which coordinates the Cu${}^{2+}$
ion, are $81.6$ kJ/mol in the holo state, $56.6$ kJ/mol in the
(Cu,E) % Zn-depleted
state, and $44.7$ kJ/mol in the (Cu-shift,E) 
state. That is, Zn-depletion destabilizes Cu-binding residue $120$,
and shift of the Cu ion to the Zn-binding position further destabilizes that residue. 
Similarly, the work values for residue $80$, which coordinates Zn${}^{2+}$, 
are $73.1$ kJ/mol in the holo state, $57.6$ kJ/mol in the (Cu,E) 
state, and  $68.6$ kJ/mol in the (Cu-shift,E) 
state. That
is, Zn-depletion destabilizes coordinating residue H$80$, and shift of
the Cu to the Zn-position, a free energetically favorable process,
partially recovers the mechanical stability of that residue in the WT protein.

The distribution of apo SOD1
is broadened compared to (Cu,E) 
SOD1, but not weakened overall,
(the mean work for both variants was within $0.2$ kJ/mol).
That is, the most weakly mechanically
stable residues in apo SOD1 are less stable than the weakest
mechanically stable residues in (Cu,E) 
SOD1 (though these residues are not the same), 
but the most stable residues of apo SOD1 are more stable than the most
stable in (Cu,E) 
SOD1 (and these residues are also not the same in the two variants).
}

\subsection{\scriptsize{Metal binding and disulfide bonding are cooperative:
  removing one destabilizes the other}}
\label{sectssbond}

\begin{figure}[H]
\includegraphics*[width=8cm]{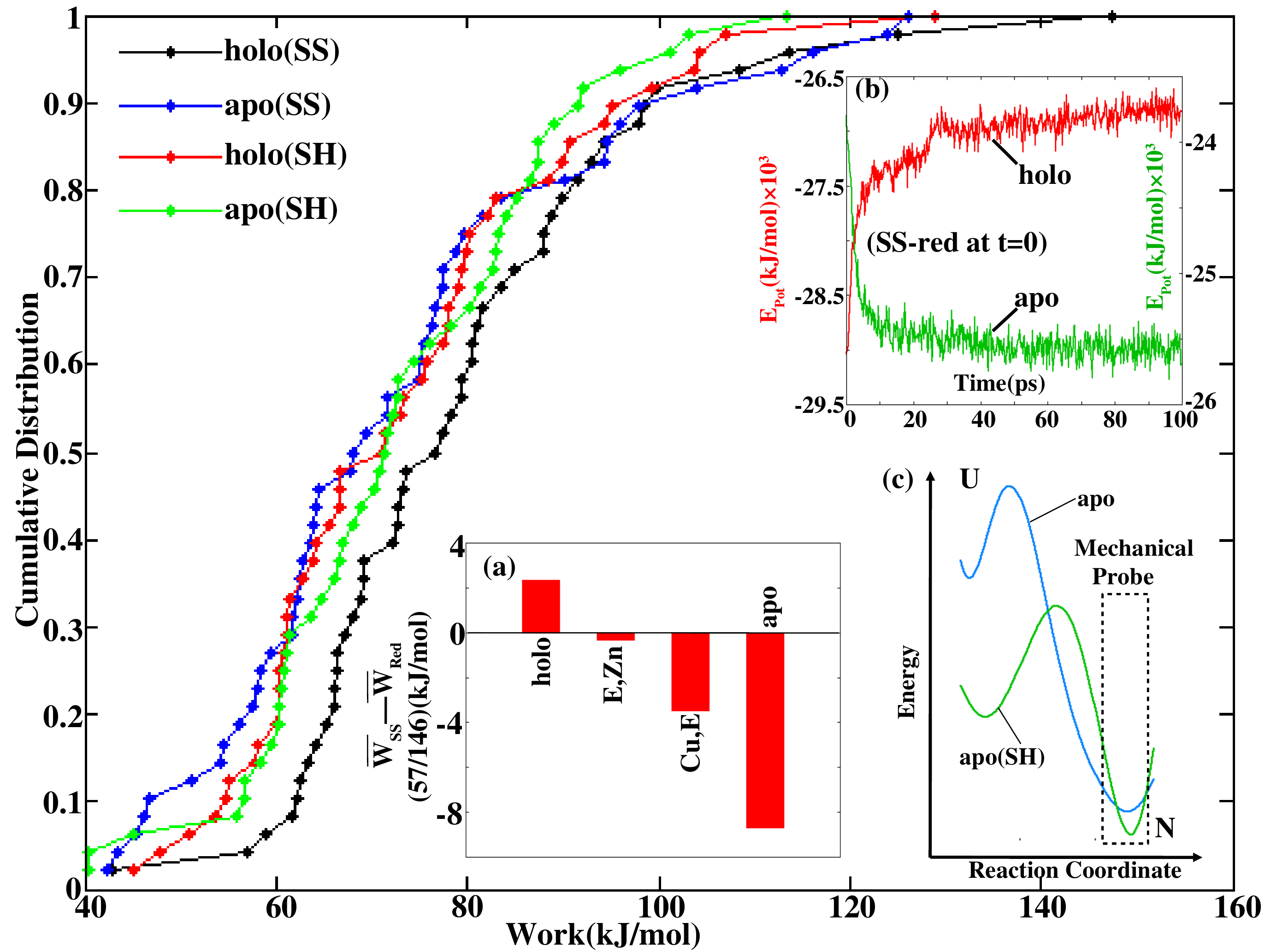}
\caption[]{
\tiny{Interplay of disulfide linkage and metal depletion on protein stability.
(Main panel) Cumulative distributions of work values for holo, apo,
holo(SH), and apo(SH) SOD1. Both metal depletion and
disulfide reduction reduce the stability of the WT protein,
but metal depletion generally results in larger destabilization for work values
less than about 80 kJ/mol.
Comparison of the holo, holo(SH), and apo(SH) cumulative
distributions indicates that the presence of the disulfide
bond stabilizes the holo form of SOD1, but destabilizes the apo form
of SOD1.
(Inset a) Bar-plot of work differences for several SOD1 variants using
the following procedure. We first found the average work, $\overline{W}$, to pull residues $57$ and $146$ to $5$ \AA,
for both disulfide-containing and disulfide-reduced forms of holo,
(E,Zn), (Cu,E) and apo (metal-depleted) SOD1. Then we
took the difference in average work between the
disulfide present and disulfide-reduced forms for all those
variants $\overline{W}_{SS} - \overline{W}_{Red}$. This was taken
as a measure of effect of disulfide bond on the stability of SOD1.
The disulfide bond stabilizes the holo form, but destabilizes
the (Cu,E) and apo forms.
(Inset b) Potential energy as a function of time after initial
reduction of the disulfide bond at $t=0$. Red curve is for holo SOD1,
green curve is for apo SOD1.  (Inset c) Schematic free energy profiles
of apo and apo(SH) SOD1 (see Discussion). The
key features of the profiles are a stiffer native basin,
a moderately lower free energy in the native state, and a
substantially lower free energy in the unfolded state, for apo(SH)
SOD1 relative to apo SOD1. The mechanical profiles probe the region
outlined in the dashed box.
}}
\label{figEffectSS}
\end{figure}

\scriptsize{Figure~\ref{figEffectSS} plots the cumulative distributions of the
mechanical work profiles for holo, apo, 
holo/SS-reduced (holo(SH)), and apo/SS-reduced (apo(SH)) SOD1. 
All modifications destabilize holo SOD1. By comparing apo with
holo(SH) SOD1 however, 
metal depletion has the largest destabilizing
effect with only a few exceptions, 
over the bottom 
77\% of the work values. The cumulative distributions then cross at
around $81$ kJ/mol. 

The weakest regions, with work values less than $60$ kJ/mol, are on
average about 5 kJ/mol weaker in apo SOD1 than in holo(SH) SOD1, while the residues with moderate
mechanical stability in the range of $60-80$ kJ/mol are only about
$1.4$ kJ/mol weaker on average for apo SOD1. In this regime, 
disulfide reduction has nearly as much destabilizing effect as metal
depletion. The most mechanically stable regions, with work values
$> 86$ kJ/mol, are on average about 5.5 kJ/mol weaker
for holo(SH) SOD1 than for apo SOD1, i.e. disulfide reduction is more
destabilizing for the most mechanically stable residues requiring the
highest work values. It is worth emphasizing again that these 
regions in the cumulative distribution need not involve the same
residues for the 2 variants. 

Perhaps surprisingly, metal depletion appears to have a moderately stabilizing
effect on holo(SH) 
SOD1, at least over the range of work
values less than 81 kJ/mol. Apo(SH) 
SOD1 contains the 3
weakest residues between the holo(SH) and apo(SH) variants, however
for $19$ out of the total of $31$ residues of the 
mechanical scan with work values less than 81 kJ/mol, apo(SH) 
SOD1 is more stable than holo(SH) SOD1. 
This is seen directly from the cumulative distributions. 
This observation prompted a study of the mechanical rigidity of SOD1
variants in the vicinity of the residues involved in the disulfide
bond, for various metallated states. For a given metallated state of the
protein, we recorded the mean mechanical work to
pull residues 57 and 146 (involved in the disulfide bond) to 5 \AA,
for both the disulfide-present and the disulfide-reduced state. If the
disulfide bond is present, both residues will tend to move together
when one is pulled. 
Inset (a) of Figure~\ref{figEffectSS} plots
the difference, SS-present minus SS-reduced, of the mean mechanical
work to pull residues 57 and 146, for holo, (E,Zn), 
(Cu,E), 
and apo SOD1. 
For holo SOD1, the presence of disulfide bond results in mechanical
stabilization of these residues, or stronger coupling to the rest of
the protein. The effect goes away once Cu is removed from the protein.
If Zn is removed, the reverse effect is observed: formation of the
disulfide bond weakens the mechanical coupling between residues 57/146
and the rest of the protein.  The mechanical weakening effect of the
disulfide bond is most significant in the apo state of the protein. 

Just as metal depletion mechanically stabilizes the disulfide-reduced state of the
protein, so also does disulfide-reduction} 

\scriptsize{stabilize the metal-depleted
protein.
Comparing the cumulative distributions of apo(SS) and apo(SH) SOD1, 32
of 38 work values less than 86 kJ/mol are more stable for apo(SH) than
for apo(SS). Thus disulfide-reduction mechanically stabilizes at least
the weaker regions of the apo protein. 

What about the effect on the overall potential energy in the protein?
We considered the following kinetics study: in both holo and
apo SOD1, we instantaneously reduced the disulfide bond, and
investigated the potential energy in the protein as a function of
time. 
Inset (b) of  Figure~\ref{figEffectSS} plots the potential energy as a
function of time immediately after disulfide reduction, for the
holo protein (red), and the apo protein (green). The
holo protein indeed shows an increase in potential energy
indicating a destabilizing effect due to disulfide reduction. However,
consistent with the above observations from the cumulative distribution
and the local work on disulfide-participating residues, the potential
energy in the apo protein decreases with
time after reduction. This indicates that the presence of the disulfide bond strains
the apo protein, and the potential energy in the protein is thus lowered by its
reduction.

The converse study of removing metals from the protein with the
disulfide bond either present or absent is contained in
Figure~\ref{figCuZnFreedimer} (a) and Table~S1. 
The holo(SH) protein has
moderately reduced affinity for
Cu and Zn, by about $5.9$ kJ/mol and $3.9$ kJ/mol respectively. 
Thus, reducing the disulfide bond lowers the affinity for the metals, and
removing the metals changes the effect of disulfide bond formation
from protein-stabilizing to protein-destabilizing. 
}

\section{\scriptsize{Discussion}}
\label{discuss}

\scriptsize{Operating on the premise that ALS-associated mutant sequences of SOD1 protein
had different mechanical properties than the WT protein, we have
undertaken a computational study to probe these mechanical
differences. By employing a combination of pulling simulations and
umbrella sampling with the weighted histogram analysis method (WHAM),
we can computationally investigate disease-relevant misfolding
processes such as loss of native structure, metal loss, and
monomerization of native dimers. We found the likelihoods of these
processes to be significantly greater in ALS-associated mutant
proteins, supporting the idea that although the consequences for
folding thermodynamic stability of many ALS-associated mutants
may be 
modest~\cite{RodriguezJ2002,HaywardL2002},
the effects on dimer
stability and metal affinity can be large. 
The commonly-used monomeric F50E/G51E mutant provides a clear precedent for
such effects. 
Moreover, dimer and metal affinities can be
influenced in subtle ways: mutations distant from the Zn or Cu can have
large effects on metal affinity, and mutations far from the dimer
interface can have large effects on dimer stability. 

Mechanical probes were employed by simulating tethers on the center
residue of SOD1, and on various residues on the protein surface.
The experimental analogue to such an {\it in silico} approach would
require multiple AFM or optical trap assays involving numerous residue
pairs about the protein surface as tethering points. This is difficult and time-consuming
to achieve in practice, which thus provides an opportunity for the present
simulation approaches.
Such a study of the native surface malleability for mutants of SOD1
may be relevant to understanding the process of intermolecular
transmission of misfolding in the cell~\cite{GradLI11}.

The work to pull a given residue to a distance sufficient to
constitute an anomalously large fluctuation, e.g. 5\AA, can be
calculated as a function of sequence index for a given SOD1
variant. This results in a characteristic mechanical profile for that
protein. We found that such a profile was 
significantly different between WT SOD1 and other
post-translationally modified (PTM) variants such as metal depleted or disulfide
reduced SOD1. A systematic comparison of mechanical profiles between
WT SOD1 and ALS-associated mutants is an interesting and important
future study. 

Some of our {\it in silico} observations are
  consistent wth previous experimental benchmarks, however we have
  also made several new observations here that are experimentally
  testable. Table~\ref{tabvalandpredict} gives a summary of the experimentally-validated
  results in this paper, as well as the main
  experimentally-testable predictions contained in the paper. We
  elaborate on the entries in this table in the discussion below.

A given PTM globally modulates the mechanical profile, inducing both
local and non-local changes in stability. These changes can be % both
destabilizing in some regions and stabilizing in others. Understanding
such stability changes well enough to predict them is a difficult
challenge, and would involve quantifying the interaction networks
throughout SOD1, and how these networks transmitted stresses when one
region of the protein was strained. 
A detailed study
of the consequences of such mutations, for example by studying the 
long-range communication through interaction networks in the
mutant vs WT protein (see for example Khare and Dokholyan~\cite{KhareSD06pnas})
is an interesting topic of future research.

The free energy to remove metals or monomerize SOD1 variants may % also
be calculated using WHAM. For these processes, harmonic distance
constraints between $C_\alpha$ atoms were applied to prevent the
inevitable structural deformations that occur in such a non-equilibrium pulling assay,
which would add to the measured free energy.  The free energy as
obtained from the Jarzynski equality (without distance constraints)
was thus always greater than the 
free energy obtained from the WHAM protocol along with harmonic
distance constraints. Removing the distance constraints from the WHAM
protocol reproduced the Jarzynski-derived free energy.

We found that many discrepancies of SOD1 variants
from WT were difficult to disentangle from the work profile, but often
emerged most naturally from the cumulative distribution of work values. 
Mechanical scans were thus used to construct the cumulative
distributions, which then allowed us to distinguish the
stabilizing energetics in various forms of PTM SOD1. 
In principle, different mechanical profiles could give rise to similar
cumulative distributions, but in practice this was never an issue. If
cumulative distributions for two variants happened to be similar, their
mechanical profiles could always be compared. 
Many of the properties we may be interested in for a given SOD1 variant involve
coarse-grained properties of the protein that can be directly obtained
from the cumulative distribution, such as how many residues are weakly
stable given a threshold of work that perturbs the system.

By comparing the cumulative distribution of the work values for holo
and metal depleted SOD1 variants, we found that loss of either metal makes
the protein more susceptible to structural deformation. Loss of Zn had
a larger effect than loss of Cu. The loss of either metal modulated
the entire mechanical profile, such that by pairwise comparison, the
work values no longer showed strong correlation with those in the
holo protein. The strongest correlation with holo protein was seen for
(Cu-shift,E) SOD1, which lacks Zn and has Cu shifted to the Zn
position ($r=0.32$, $p=0.03$).} 

\renewcommand{\arraystretch}{1}
\begin{table*}
\scriptsize
\caption{Experimental validation and predictions of the model}
\begin{tabular}{l|c|c}
%\hline
\multicolumn{3}{c}{ } \\
{\bf Property} & {\bf Simulation} & {\bf Experimental Reference} \\
\hline
$\Delta G$ of apo WT dimer $\rightarrow$ apo monomer & 62 kJ/mol (Fig. 3b) & 61 kJ/mol~\cite{LindbergMJ2004:pnas}${}^a$ \\
\hline
$\Delta G$ of apo G85R dimer $\rightarrow$ apo monomer & 47.9 kJ/mol
(Fig. 3b) & 48.9  kJ/mol~\cite{VassallKA06} \\
\hline
Non-local destabilization of & \multirow{2}{*}{Fig. 3b (and text)} & \multirow{2}{*}{~\cite{VassallKA06},~\cite{KhareSD06pnas}${}^b$} \\
dimer stability of SOD1 mutants & & \\
\hline
\multirow{2}{*}{Zn-binding $\Delta G$ for WT SOD1} &  \multirow{2}{*}{40 kJ/mol (Fig. 3a)} & 40 kJ/mol~\cite{KhareS2004}, 45 kJ/mol~\cite{PotterSZ07}${}^c$, \\
 & & 56 kJ/mol~\cite{KayatekinC2010}, 76 kJ/mol~\cite{CrowJP97} \\
\hline
Cu-binding $\Delta G$ for WT SOD1 & 53 kJ/mol (Fig. 3a) & 98 kJ/mol~\cite{CrowJP97} \\
\hline
Zn-binding $\Delta \Delta G$ for mutant SOD1 & 3.2(A4V), 4.6(G93A), & -2.9(A4V), -1.7(G93A), \\
(value(mutant)) r=0.99,p=0.01 & 7.7(L38V), 12.3(S134N) kJ/mol & 2.9(L38V), 6.7(S134N) kJ/mol~\cite{KayatekinC2010} \\
\hline
Cu-binding $\Delta \Delta G$ for mutant SOD1 & 8.1(A4V), 14.9(I113T), & 1.2(A4V), 2.1(I113T), \\
(value(mutant)) r=0.78${}^d$ & 12.0(l38V) kJ/mol & 2.4(L38V) kJ/mol~\cite{CrowJP97}\\
\hline
Cu,E SOD1 is more malleable than E,Zn SOD1 & \multirow{3}{*}{Fig. 5} & \multirow{3}{*}{~\cite{ElamJS03nsb,RotilioG72,LippardSJ77,KayatekinC08,NordlundA2009}} \\
(Zn-binding facilitates native & & \\
structure moreso than Cu binding) & & \\
\hline
Zn-binding and electrostatic loops & \multirow{2}{*}{Fig. S7} & \multirow{2}{*}{~\cite{RobertsBR07,BanciL03,StrangeRW03,ElamJS03nsb,NordlundA2009}} \\
are selectively destabilized by metal release & & \\
\hline
apo(SS) is less stable than holo(SH) SOD1 & Fig. 6, Fig. S6, Table S5 & ~\cite{FurukawaY05} \\
\hline
Validation of Jarzynski equality with & \multirow{2}{*}{Fig. 2} & \multirow{2}{*}{~\cite{GoreJ03}} \\
WHAM for native mechanical stability & & \\
\hline
\multicolumn{3}{c}{ } \\
%\hline
{\bf Predicted Property} & {\bf Simulation} & {\bf Suggested Experiment} \\
\hline
Dimer stability for SOD1 variants and ALS & \multirow{3}{*}{Fig. 3b} & \multirow{3}{*}{Equilibrium denaturation} \\
mutants${}^e$ (all had reduced $\Delta G$ w.r.t. WT; & & \\
D124V, H80R, and G85R were among lowest) & & \\
\hline
Zn-binding/Cu-binding $\Delta G$ for SOD1 variants  & \multirow{4}{*}{Fig. 3a} & \\
and ALS mutants${}^f$, (all had reduced $\Delta G$ & & Chelation assay, Inductively coupled \\
w.r.t. WT; T54R was among lowest; D125H & & plasma mass spectroscopy (ICP-MS) \\
has comparable Cu and Zn affinities) & & \\
\hline
Correlation between Zn-binding & Fig. 3a and corresponding text &  ICP-MS, Chelation assays \\
and Cu-binding $\Delta G$ values & r=0.93,p=4\mbox{e-}10 & (In~\cite{HaywardL2002}, r=0.87(0.84),p=5\mbox{e-}6(1\mbox{e-}4)${}^g$) \\
\hline
Quantitative valence dependence of Cu & \multirow{2}{*}{Fig. 3a} & Competitive chelation \\
affinity; Cu${}^{1+}$ has lower affinity than Zn & & (see e.g.~\cite{FeagaHA11}) \\
\hline
Larger effect of Zn on Cu & \multirow{2}{*}{Fig. 3a} & ICP-MS, \\
affinity than Cu on Zn affinity &  & Chelation assays \\
\hline
The putative Zn-binding site is more & \multirow{4}{*}{Section 2.4} & EXAFS; Relative occupation \\
favorable for Cu (Cu-binding is under & & of Cu- and Zn-binding sites in \\
kinetic control); Kinetic barrier & & crystal structures in absence of Zn\\
separates binding sites & & \\
\hline
Shift of Cu to Zn-binding site stabilizes & \multirow{2}{*}{Fig. 5} & AFM/Optical trap or DSC of \\
SOD1 relative to Cu,E SOD1 & & E,Cu SOD1, derived perhaps by chelation \\
\hline
Cooperativity between metal binding & \multirow{2}{*}{Fig. 6}  & AFM/Optical trap with \\
and disulfide formation & & Chelator(DTPA)/Reductant(TCEP)\\
\hline
Strain/frustration in the apo state & Fig. 6, Table S5 (holo/apo) & Relaxation-dispersion NMR, H/D exchange \\
\hline
Non-local effects of mutations or & \multirow{2}{*}{Fig. 5, Fig. S1} & Multiple-tethering-point AFM \\
metalation state on mechanical stability & & or optical trap probes of SOD1 \\
\hline
Mechanical stabilization of apo(SH) SOD1 & \multirow{2}{*}{Fig. 6} & \multirow{2}{*}{AFM/Optical trap${}^h$} \\
with respect to apo(SS) SOD1 & & \\
\hline
Cu,E(SS) SOD1 is marginally less & \multirow{2}{*}{Fig. S6, Table S5} & DSC, Equilibrium denaturation; \\
(mechanically) stable than holo(SH) SOD1 & & Chevron kinetics
(AFM/Optical trap) \\
\hline
\end{tabular}
\\
${}^a$Dimer binding free energies for pseudo-WT C6A/C111S are
somewhat less, about
$51$-$52$kJ/mol/dimer~\cite{LindbergMJ2004:pnas,VassallKA06,SvenssonA-K2010}. \\
${}^b$Simulation result.  \\
${}^c$This number is reported as an approximate lower bound to the affinity.  \\
${}^d$Dataset is too small (N=3) for the correlation to have statistical significance. \\
${}^e$Mutants with newly predicted dimer binding free energies:
holo(SH), (Cu-shift,E)(SS), Cu,E(SS), E,Zn(SS), E,Zn(SH), the
following apo mutants: T54R, H80R, and G127X, and the
following holo mutants: A4V, G37R, L38V, G41D, G41S, H43R, H46R,
H46R/H48Q, T54R, D76Y, H80R, G93C, I113T, D124V, D125H,
G127X, S134N, L144F.  \\
${}^f$Mutants and variants with predicted metal binding free energies for Zn: holo(SH), E,Zn(SS), E,Zn(SH), W32S, G37R, G41D, G41S, H43R,
H46R, H46R/H48Q, T54R, D76Y, H80R, G85R, G93C, D124V, D125H, G127X, L144F; for Cu: holo(SH), (Cu-shift,E)(SS), Cu,E(SS), W32S, G37R, G41D, G41S,
H43R, H46R, H46R/H48Q, T54R, D76Y, H80R, G85R, G93A, G93C, D124V, D125H, G127X, S134N, L144F. \\
${}^g$Proteins used: WT, A4V, L38V, G41S, G72S, D76Y, D90A, G93A, E133$\Delta$, H46R, H48Q, G85R, D124V, D125H, S134N (numbers in parentheses
do not include H46R, H48Q and D125H). \\
${}^h$The fact that NMR measurements report no significant change in tertiary structure upon disulfide reduction of monomeric apo(SS)
SOD1~\cite{ArnesanoF04} is consistent with this prediction.
\label{tabvalandpredict}
\end{table*}

\scriptsize{Interestingly, the work values of apo
SOD1 showed modest anticorrelation with those in holo SOD1 ($r=-0.37$,
$p=0.01$), consistent with the idea that mechanically rigid residues
in the holo protein are stabilized by the presence of metals, and that
mechanically rigid residues in the apo protein tend to be destabilized
by the metals. This competition between apo-state stability and metal
affinity-driven holo-state stability was also seen in
mutant studies, which revealed significant strain in the apo
state~\cite{DasA12pnas1}. 

Our simulations showed that (Cu,E) SOD1 is substantially more mechanically malleable than
(E,Zn) SOD1, which is consistent with previous experimental data: Early work by
Rotilio {\it et al} found that the extent of recovery of the native
copper site in terms of spectroscopic features and catalytic
activity was related to the amount of residual zinc present~\cite{RotilioG72}, implying
that Zn aided the incorporation of Cu. Work from the Valentine group
later showed, by analyzing NMR shifts of coordinating histidine residues in the
presence of a chemical modifier contingent on solvent exposure~\cite{LippardSJ77},
that Zn aided the incorporation of Cu by structuring the protein, and
that Zn alone was sufficient to structure the protein. 
Matthews and co-workers found substantial stabilizing free energies
upon Zn binding, for all phases of structural organization of SOD1,
dimeric to unfolded~\cite{KayatekinC08}.

Crystal structures of mutants that either inadequately or anomalously
bind Zn show amyloid filament-like non-native interactions, further
supporting the critical structural role of Zn.  Elam et
al~\cite{ElamJS03nsb} find that the crystal structures of apo H46R,
and S134N, for which Chain B has only 1 Zn ion bound in the Cu
position, both show similar and significant structural disorder in
which non-native contacts are made between the electrostatic loop from
one domain and an exposed cleft made by $\beta$ strands 5 and 6 of a
neighboring domain.  Oliveberg and co-workers find a similar effect
for the H63S/H71S/H80S/D83S variant~\cite{NordlundA2009}, which cannot
bind Zn in the putative binding site, but instead binds Zn in the Cu
binding site for about 3/4 of the molecules.  Histidine 48 fails to
ligate the Zn ion and instead swings outwards, leading to structural
disorder in loop IV, which in turn destabilizes Loop VII. This loop
then interacts with edge $\beta$ strands in neighboring molecules,
analogously to the apo H46R and S134N crystal structures.

Consistent with the observations that the role of Zn is structural
while the role of Cu is enzymatic, we found that the affinity for Cu
is reduced by $\approx 11.4$~kJ/mol in the absence of Zn, 
while the affinity for Zn is reduced to a lesser degree by $\approx 5.9$~kJ/mol 
in the absence of
Cu. In the simulation protocol we employed, all structures were
equilibrated before applying constraints between C$\alpha$ atoms, so
the structural deformations associated with loss of the partner metal prior to
the affinity measurement were accounted for. 
As well, we also accounted for the effects due to re-equilibration of the protein
structure in the final state of the protein after extraction of the 
metal, in calculating the net affinity. 
This relaxation effect is more significant for Zn than for Cu due to
the stronger coupling of native structure to Zn binding, and most
significant for dimer separation, because of the removal of steric
constraints and large change in interaction energies (see Figures~S3,
S4, and~S5). 
However, metal binding free energies were obtained from a metal expulsion
process along ``the most likely'' pathway, but they did not consider
all possible pathways of escape- the contribution of all other
pathways is an entropy contribution that would likely result in modest
reduction in the estimate for the binding free energy.
Nevertheless, our
measurements should give a good approximation to the relative
differences in affinity between species. The decreased metal
affinity of ALS-associated mutants points towards 
the general role of metal loss in the misfolding and propagation
process.

Matthews and co-workers obtain larger values of 
Zn binding free energies of about
$56$
kJ/mol/monomer~\cite{KayatekinC2010}. Although these values are obtained from several
approximations, including the use of C6A/C111S and C6A/C111S/F50E/G51E
as reference states for dimeric and monomeric WT SOD1, it is
instructive to investigate reasons on the computational side as to why
their value might be larger than our value of $40$ kJ/mol. Our calculations
obtain the binding free energy for monomeric SOD1, and so neglect
cooperativity effects between monomers in Zn binding to the dimer
state. As well, although the OPLS-AA/L force field used in our
simulations attempts to account for the electronic polarization of
Histidines which are critical in modulating metal binding free
energies, polarization may still be underestimated by the force field.
The discrepancy between our value of Cu-binding free energy to WT
  SOD1 and that of Beckman and colleagues~\cite{CrowJP97} indicates
  that Histidine polarization may be particularly important in Cu-binding.

Further information on the loss of structure due to metal depletion is
obtained from studies of the solution structure 
of the obligate monomer mutant E133Q/F50E/G51E, for which 
Banci {\it et al.}~\cite{BanciL03} found that the electrostatic and zinc-binding loops
were severely disordered and extensively mobile in the absence of both
metals, but not in the absence of Cu alone, i.e. for the (E,Zn)
variant. Crystal structural studies of the apo WT protein also confirm
that for monomers devoid of metals,  the electrostatic and Zn-binding
loops are disordered 
and not visible in the electron density maps~\cite{StrangeRW03}. Some
monomers in this study with 20\% occupancy at the Zn site possessed well-ordered
Zn-binding and electrostatic loop regions, however this was thought to
be due to crystal packing forces. 
Beckman and colleagues have crystallized a
constitutively Zn-deficient but Cu-containing SOD mutant
H80S/D83S/C6A/C111S by mutating 
two zinc-binding ligands to hydrogen-bonding
serines~\cite{RobertsBR07}. This variant was observed to have a
disrupted dimer interface and partly disordered zinc binding and
electrostatic loops. 
Both our mechanical pulling and simulation results are consistent
  with these observations (Table~\ref{tabvalandpredict}). 
In Figures~S7b and c, the increase in mechanical malleability and
  magnitude of fluctuations upon metal release are measured for several regions SOD1,
  including residues 1-48, residues 48-83 constituting the Zn-binding
  loop, residues 84-120, residues 121-142 constituting the
  electrostatic loop, and residues 143-153. These plots show that
  loops IV and VII undergo systematic softening and increased
  fluctuations compared to other regions, as metals and in particular Zn are lost from the
  protein.

The mechanical response to the loss of metals that we observed was not localized to loops IV
and VII, and in this sense apparently differs from the above
observations and constitutes a prediction of the model
(Table~\ref{tabvalandpredict}). For 
example, while the mean mechanical stability of loops IV and VII
decreased substantially after the loss of ZN or both metals, 
some regions of the protein such as $\beta$ strands 3 and 6 are more
rigid for the apo and (Cu,E) proteins than for the holo
protein. 
Metal depletion of the WT enzyme modulated the entire
mechanical stability profile, locally and nonlocally. 

Our pulling simulations, which incur a significant  perturbation on
the structure, measure a quantity that is distinct from thermal
fluctuations in the native ensemble. 
As such, they may show discrepancies with experimental probes
  measuring equilibrium fluctuations. 
Supporting this, the work values to extend to 5\AA~do not correlate
with the RMSD values for the
corresponding residues (Table S3, Supplementary Content). 
As well, normal modes of the fluctuation spectrum for
beta sheets may involve, for example,
sliding motions that are transverse to the radially-directed pulling
simulations that we employed. On the other hand, misfolding-related
processes likely involve motions that are larger deformations beyond native basin fluctuations, so
the information gained from relatively large mechanical perturbations can be
useful in understanding misfolding.

We saw that post-translational modifications, as well as ALS-associated
mutations, reduced both metal affinity and dimer stability. This is
consistent with experimental observations that mutations seemed to
favor increased formation of Zn-free monomeric
intermediates~\cite{RumfeldtJA09}. 
Several ALS-associated mutations that we investigated reduced
holo dimer stability moreso than did disulfide
reduction or loss of metals from the holo WT protein. 
On the other hand, apo(SH) WT SOD1 had lower dimer stability than
  15/22 holo ALS-associated mutants; metal depletion
along with disulfide reduction has larger consequence for monomerization
  than many mutations do alone. The apo mutants may have lower dimer stability
  still, however. 
The decrease in dimer stability of holo WT protein due to disulfide
reduction, about $37$~kJ/mol,  is more significant than depletion of
either metal, but not both (about $43$~kJ/mol). 
The decrease in dimer stability due 
to Cu depletion is the smallest change of all modifications we
investigated, about $12$~kJ/mol, while the decrease in dimer stability
due to Zn depletion is significant: about $28$~kJ/mol. 
Increased
concentrations of oxidatively-modified monomeric species have been observed on pathway to 
aggregate formation by dynamic light scattering
measurements~\cite{RakhitR04}, supporting the notion of monomers as an
amyloid precursor. As well, aggregation rates have been
observed to increase with decreased concentration of SOD1, which
strongly indicates
that monomeric species act as precursors to aggregation~\cite{KhareS2004}.
Small molecules targeted to stabilize the dimer interface have been observed to
significantly inhibit {\it in vitro} aggregation 
in mutants A4V, G93A, G85R~\cite{RaySS05}.

We found the dimer binding free energy of apo WT protein to be about $62$~kJ/mol, 
which is in quite good agreement with experimental estimates derived
from equilibrium denaturation data
on
C6A/C111A double mutants by extrapolating the
destabilizing effects of the C6A/C111S double mutation
($\Delta G_{2M\rightarrow M_2} \approx 61$~kJ/mol~\cite{LindbergMJ2004:pnas}, see
Table~\ref{tabvalandpredict}). 
Fits of equilibrium 
denaturation data of the pseudo-WT mutant protein C6A/C111S 
to three-state models give somewhat smaller numbers
for dimer stability for the double mutant of about
$51$-$52$~kJ/mol~\cite{VassallKA06,SvenssonA-K2010,LindbergMJ2004:pnas}.
Desideri and colleagues~\cite{StroppoloME00} report a dimer stability
for apo WT protein of about $52$~kJ/mol, however
these studies also predicted moderately smaller dimer affinity for the holo
protein than the apo protein (about $49$~kJ/mol), in contrast to
  our results. 

The ALS-associated mutant S134N has been observed experimentally to have only moderate
effects on thermodynamic stability~\cite{KayatekinC2010}, indicating
mechanisms aside from global unfolding play a role in disease
propagation. Indeed, 
we observe significant effects on both metal affinity and dimer
stability for this variant: the dimer stability is reduced from the WT
by $45$~kJ/mol, and the Zn binding free energy is reduced from the WT
by $12$~kJ/mol. Matthews and co-workers also see a Zn binding free
energy for an obligate monomeric form of this mutant
(C6A/C111S/F50E/G51E/S134N) that is reduced from the
C6A/C11S/F50E/F51E form by $7$~kJ/mol, based on the ratio of their $K_d$
values~\cite{KayatekinC2010}. On the other hand, 
some of their obligate monomeric mutants have higher free energy for
Zn binding (e.g. their $\Delta \Delta G$ values for A4V, G93A, L38V,
and S134N are -2.9, -1.7, +2.9, and +6.7~kJ/mol respectively). This
contrasts with the consistent decrease in Zn affinity for ALS mutants that we have
seen here, and may be due to the mutations present in the their
pseudo-WT background. That said, our values of $\Delta \Delta G$ for Zn-binding
free energy correlate
very well with theirs (see Table~S1): $r=0.99$, $P_r = 0.01$.
Consistent with Crow et. al.~\cite{CrowJP97}, we do see reduced metal affinity
for all fALS mutants considered, however our values for Zn binding free
energy did not correlate with their values (of A4V, L38V, and I113T),
possibly due to monomerization during the course of the affinity
measurements~\cite{KayatekinC2010}. Our values for Cu binding free
energy did correlate with their values ($r=0.78$), however the number
of mutants (3) is not enough to be statistically significant.

The binding free energy of Zn to WT SOD1 monomer as measured
experimentally by Dokholyan
and co-workers is about 40~kJ/mol~\cite{KhareS2004}, which is in
excellent agreement with our value of $40$~kJ/mol. However, their value was obtained
at pH 3.5, which would reduce the value from that at pH 7. 
Likewise, calorimetry measurements at pH 5.5 by Valentine and colleagues give
  comparable numbers of at least $45$~kJ/mol for binding the first Zn
  ion to apo SOD1 homodimer~\cite{PotterSZ07}.
In~\cite{KhareS2004}, these authors also
observed that metal depletion, facilitated by dialysis against
metal-free buffer, promoted aggregation at 30$\mu$M and 
pH 3.6.  We cannot address whether metal depletion occurs before or
after monomerization along the aggregation pathway, and this does not
appear to be a settled issue. As well, the total free energy
change from holo dimer to a monomerized, (Cu,E) form is
independent of the sequence in which these changes occur, so in
principle both parallel pathways may be undertaken and would give the same
net result, provided that that part of the misfolding/aggregation
reaction is under thermodynamic control.

Our calculations for dimer stability have 14 fALS mutants in common with
the calculations of Khare et al~\cite{KhareS2006}, however we see no
correlation between our apo dimer stability values for these mutants
(Table S1) and their values for either the free 
energy of dissociation of the dimer, or the overall thermodynamic stability of
the dimer ($r=0.001$ and $r=0.08$). These authors mention that one
possible source of discrepancy between their values and
experimentally-determined stabilities is that 
their simulations relied on the assumption that there is no major
structural change in the protein upon mutation. Their values also rely on calculations
of absolute free energies~\cite{VorobjevYN01,VorobjevYN02:addendum},
for which it is often challenging to obtain 
quantitative accuracy, in contrast to relative changes in the free
energy.
We have sought to be careful here in calculating relative changes in
free energy, including relaxation after mutation or PTM by thermal
equilibration, 
and also re-equilibration in the final state after 
dimerization or metal extraction. 
Relaxation free energies in demetallation generally vary mutant
to mutant, and were more substantial for
re-equilibration of disulfide-reduced  or Zn-depleted 
SOD1 (Figures S3,S4). Relaxation free energies for monomerization were
larger than for demetallation, and were most significant for 
apo(SH) SOD1, followed by apo(SS) and Cu,E SOD1 (Figure S5). 
These variants had significantly increased entropy in the monomeric
state from
the dimeric state because of the loss of dimer interface constraints
on the zinc-binding loop.

Meiering and co-workers have found that for 
fALS mutants E100G, G93A, G85R inserted into the pseudo-WT background
C6A/C111S, dimer stability was largely unaffected by
mutation~\cite{RumfeldtJAO06}. In contrast we found significant
weakening of dimer stability due to the mutant G85R, albeit in the pure WT
background of holo protein. 
Similar investigations from the Meiering group have shown that the
same mutations in the apo dimer did weaken dimer stability, however
the decreases in dimer stability that we have calculated are
substantially larger than their values~\cite{VassallKA06}. 
These difference may be reconciled by 
computational studies of the binding free energy for fALS mutants with explicit
pseudo-WT backgrounds (e.g. triple mutants here); mutation of the
cysteines may modulate the dimer stability, and alter free energetic differences
between mutants and WT. Experimental evidence suggests however that these effects
may be modest~\cite{LindbergMJ2004:pnas,LepockJR90}, however the
computational study is an interesting topic for future work.

Several groups, including those of
Oliveberg~\cite{LindbergMJ2004:pnas}, O'Halloran and
Bertini~\cite{ArnesanoF04}, and Hart and
Valentine~\cite{DoucettePA04}, have found that
for pseudo WT
SOD, either Zn binding or formation of the C57-C146 disulfide bond
was sufficient to induce or rescue dimer formation, i.e. both the absence of 
metals (apo) and reduction or ablation of the disulfide bond are
necessary to induce monomer formation, even 
at low concentrations ($\sim 10\mu M$ SOD1). 
Sedimentation velocity analysis~\cite{DoucettePA04} indicates that both holo C57S and
apo-SOD1 both sediment as dimers, with apo-SOD1 marginally more
monomer-like. 
Consistent with these observations, we found that both metal depletion
and disulfide reduction had significant destabilizing effects on dimer
stability.
The holo(SH) dimer had lower binding free energy
than that for either the (Cu,E) or (E,Zn) variants, but
the apo dimer had lower dimer binding free energy than the holo(SH)
dimer; this result that 
binding of both metals (holo {\it vs.} apo) plays a larger role than disulfide formation in dimer
affinity is consistent with simulations from Dokholyan's group, based
on interdomain native contact data~\cite{DingF08pnas}. 
Demetallation
destabilized the disulfide-reduced dimer by about 
7 kJ/mol. However, 
we found that disulfide reduction only further destabilized the apo dimer
by about 1 kJ/mol, in apparent contrast to the observation that
apo(SS) SOD1 elutes as a
dimer but apo(SH) SOD1 elutes as a monomer. One possible resolution
here is the role that increased entropy might play in the reduced
protein in rebinding a partially disrupted interface. Further
work will be required on the computational side to reconcile these
observations.

The mutants H46R, H46R/H48Q, and D125H were atypical in that the Zn
affinity was larger than the Cu affinity (though only marginally for
D125H). The first two of these variants have 
mutations in Cu-coordinating residues, so the reversal in affinity was
not surprising. For D125H the reasons are more subtle however.
The center of mass of the
aspartic acid side chain is $12$\AA~from the Cu and $11$\AA~from the
Zn respectively, so the coupling to binding free energy involves bridging interactions
through at least 2 side chains. This system provides another example
of non-local communication in the protein through stress-strain networks.

Cu${}^{1+}$ affinity is quite modest for mutants, indicating ready
release of monovalent Cu in the reduced phase of the catalytic
cycle. Released or exposed Cu is capable of reacting nonspecifically
with a variety of substrates to produce reactive oxygen and nitrogen
species that are toxic to intracellular structures,
including microtubules, metabolic enzymes, and signalling
proteins~\cite{AhmedMS00}. Reactive species may oxidatively modify side
chains to inactivate SOD1~\cite{AlvarezB04} and induce its
misfolding~\cite{RakhitR02}; positive feedback loops have been
identified between protein misfolding and
excitotoxicity~\cite{LiptonSA07}. 

By simulating AFM-like probes, 
we investigated the mechanical stability of holo, holo(SH), apo, and
apo(SH) variants of WT protein. 
Here we found that the apo variant was less mechanically
stable than holo(SH) (statistical significance $p\approx 3\times
10^{-8}$), 
indicating that metal depletion is more mechanically destabilizing than
disulfide reduction. This is consistent with differential scanning
calorimetry (DSC) measurements of the melting temperatures of SOD1 variants,
where (E,Zn)(SH) SOD1 was observed to have larger $T_m$ than the apo % E,E/SS
variant by about $9{}^\circ$C~\cite{FurukawaY05}. 
Similarly, Cu,E(SS) is more mechanically stable than holo(SH) SOD1
(Figure~S6), and we anticipate that DSC and other measurements of this
system would recapitulate the results of apo(SS) and holo(SH) (Table 2).
Intriguingly, we found that apo(SH) monomer was more mechanically stable
than apo monomer ($p\approx 8\times 10^{-8}$). 
This observation was supported by time-resolved simulations of
the relaxation kinetics of the internal potential energy. 
These showed that the protein internal potential
energy increased upon disulfide reduction for the holo protein,
indicating decreased stability through loss of stabilizing
interactions. However,  the protein internal potential
energy decreased upon disulfide reduction for the apo protein,
indicating {\it increased} stability and relaxation of frustrating
interactions. The existence of frustration in the apo state is
supported by analysis of the mechanical stability of C-terminal
truncation mutants, native basin dynamic fluctuation
analysis, and frustration/potential energy
analysis~\cite{DasA12pnas1}. The potential energy 
increase after SS-reduction in the holo protein indicates that entropy
increase plays a role in the lowering of the free energy after the
removal of constraints. 
The potential energy decrease after
SS-reduction in the apo protein indicates that 
the release of native stress upon removal
of constraints is sufficiently large to be observed amidst the concomitant effects arising
from entropy increase.

These results, at first sight, contrast with DSC measurements
indicating that melting temperatures are larger for WT apo SOD1
than for apo(SH) SOD1 by about
$7{}^\circ$C~\cite{FurukawaY05}, i.e. the disulfide bond
thermodynamically stabilizes the apo monomer. 
One important difference between the simulation 
probes and experimental measurements is that our mechanical probes
measure stiffness in the native basin, but not the relative difference
in the free energies between the folded and unfolded states. While
these generally correlate for one variant at different temperatures,
they need not correlate for different protein variants at the same
temperature, in particular if one variant has larger entropy in the
unfolded state, as is the case here because of the presence or
  absence of a disulfide constraint. It is possible to observe a stiffer native basin for a
less thermo-stable protein, as indicated schematically in inset (c) of
Figure~\ref{figEffectSS}. 

Zinc removal dominates the effects of metal depletion on
mechanical stability. In fact depletion of both metals results in lower
mechanical stability than (Cu,E) SOD1 only for a few ($\sim 8$) of the least
stable residues (i.e. the statistical significance that the
  cumulative distribution of apo is weaker than (Cu,E) is about
  $p\approx 9\times 10^{-3}$). 

Insertion of metals 
into apo(SH) SOD1 indeed makes some regions very mechanically
stable, increasing the stability of the most stable regions of the
protein.  However the general trend over most of the residues is to
mechanically destabilize the protein due to added internal stresses. Thus, metal
depletion modestly stabilizes the disulfide-reduced protein mechanically
($p\approx 8\times 10^{-3}$). 
Together with the observation that disulfide reduction
  mechanically stabilizes apo SOD1, this evidence points towards
  cooperativity between metal binding and disulfide formation
  (Table~\ref{tabvalandpredict}).
As mentioned above, mechanical
stability need not be equivalent with thermodynamic stability:
thermodynamic stability 
accounts for the free energy of the unfolded state, while our
mechanical stability measurements do not.

For the Zinc-depleted form, shifting of the Cu to the Zn position
resulted in a more robust mechanical stability profile ($p
\approx 1\times10^{-7}$). 
Supporting this trend in stability, WHAM simulations of the transfer
of Cu from the putative Cu binding site to the Zn binding site for the
(Cu,E) form of the protein lowered the free energy by $\approx
-6$~kJ/mol, indicating that the Zn binding position is 
more favorable for the Cu when only Cu is bound.
This provides further evidence that binding of Cu is under kinetic
control, modulated essentially by the initial Zn-coordinated folding of the
protein and its subsequent binding to the co-expressed Cu chaperone
CCS, which selectively loads Cu to its putative site~\cite{LambAL01}.
Computing the transfer free energy for Zn from the Zn-binding site 
to the Cu-binding site gives $\approx +6$~kJ/mol, supporting the model
that Zn-binding is thermodynamically controlled. Kinetics
  experiments by Oliveberg and colleagues indicate Zn$^{2+}$ binds to
  the putative Cu site during folding, but may be transferred
  to the higher affinity Zn-binding site late in the folding process~\cite{LeinartaitL10}.
We note that the above transfer free energies are estimates, and
  do not take into account the differential affinity due to the
  3d${}^{10}$ {\it vs.} 3d${}^{9}$ electronic  valence of 
  Zn$^{2+}$  and Cu$^{2+}$ respectively.

The decrease in free energy corresponding to the shift of the Cu is
accompanied by increased mechanical rigidity in the Zn binding
loop. However, some residues, such as residue 24, 75 and 120, are
significantly weakened by the shift. Consistent with significantly weakened
mechanical rigidity, the root mean squared fluctuation (RMSF) values
are increased for those residues, corresponding to an increase in
local entropy. Similar effects were seen here for processes such as metal
binding, and disulfide formation- energetically favorable processes
that result in local enhancement of structure, but which may increase
entropy non-locally.  Non-local entropy generation has also been seen in
pulling simulations, where local mechanical stress induced by
pulling on a particular residue resulted in non-local unwinding of the
short stretches of $\alpha$ helix in the Zn-binding and electrostatic
loops~\cite{DasA12pnas1}. 

These effects perhaps foreshadow one of the
essential features of disease propagation in template-directed
misfolding, namely that of non-local entropy transduction across the
protein. It is feasible that the induced misfolding of a protein due
to the interaction with  misfolded template would correspond to low
enthalpy and low entropy in the vicinity of the binding interface, and
high enthalpy and high entropy away from the binding interface. Such a
phenomenon would facilitate the 
structural plasticity required to render the protein amenable to
adopt altered conformations in the presence of misfolded
template. Once altered conformations are adopted, the
misfolded protein becomes part of the infectious species, capable of further
seeding misfolding through the reservoir of healthy protein. 
}

\section{\scriptsize{Methods}}
\label{methods}

\subsection{\scriptsize{Steered molecular dynamics simulations}}
\label{sectsteeredmd}

\scriptsize{Steered molecular dynamics 
(SMD) with constant-velocity moving restraints involving two tethering
points was used to simulate the action of a moving AFM cantilever on a protein. 
Three computational assays were investigated: \\
{\bf 1.}  A study of SOD1 monomer, where one tether was placed at the position
of the alpha carbon closest to the center of mass of the protein
(generally $C_\alpha (46)$), and another tether was placed on the
alpha carbon of a particular amino acid to be pulled on.     \\
{\bf 2.} A study of monomerization from the dimer, with a tether on the alpha carbon atom closest to center of mass of each monomer. \\
{\bf 3.} A study of metal removal, with one tether on the C$_\alpha$ atom of either residue Phe45 (for Cu
removal) or that of Asp83 (for Zn removal), and the other tether on
either the Cu or Zn metal ion. These residues are chosen because they
determine the approximate direction of highest solvent exposure of the metal.

In assay (1), to implement a mechanical stability scan across the surface of the
protein, every 5th amino acid along with the first and last residues was
selected for a pulling simulation. This coarse-grains the mechanical
profile, so that the work values obtained are a sampling of the
work values for all residues. 

As described in~\cite{DasA12pnas1}, 
SMD
simulations~\cite{SotomayorM2007,Carrion-VazquezM2003,IrbackA2005,ImparatoA2008}
were preformed in GROMACS using an all-atom representation of the
protein, with OPLS-AA/L parameters~\cite{JorgensenWL96,KaminskiG2001},
and a generalized Born surface area (GBSA) implicit solvent model with
protein dielectric constant 4, solvent dielectric constant 80, and 
surface tension coefficient for solvation free energies of 0.005
kcal/mol/(\AA$^2$)~\cite{BjelkmarP2010}.  
Born radii are calculated from the Onufriev-Bashford-Case
(OBC) algorithm~\cite{OnufrievA2004}, and bond lengths containing a
hydrogen atom are constrained
by the LINCS algorithm~\cite{HessB1997}.
The long-distance cut-off used for non-bonded interactions was 14 \AA~ for
both Electrostatic and van der Waals (VDW) interactions.

After energy minimization by steepest descent to remove potential
steric clashes, simulations were implemented using an integration
time-step of 2 fs, in a cubic
simulation box with periodic boundary conditions and size
such that all protein atoms were initially at least 20 \AA~ from any
cubic face.  An equilibrium simulation
was always run for 20 ns to equilibrate the protein before any pulling
simulations were performed and data was collected; coordinates were
saved every 100 ps.  

Because only moderate deformations were required to construct
mechanical force-extension curves, a very slow pulling speed (for
simulations) of $2.5\times 10^{-3}$m/s was admissible when work values
were computed. This speed is $400-8000\times$ slower than the speeds
generally used in simulations~\cite{LemkulJ2010,KimT2006,GeriniM2003},
and is comparable to pulling speeds used in AFM
experiments~\cite{LvS2010}.
The spring constant of the simulated AFM cantilever was taken to be
$5$ kJ/mol/\AA$^2$, 
which is comparable to the protein effective spring constant.%  is comparable to

For assay (1), each pulling simulation ran until the change in
distance between the two tethering points was 5 \AA.    
For assays 2 and 3, the simulations were continued until the force
fell to a value of zero indicating the separation of the dimer or demetallation of
the monomer, respectively.   
}

\subsection{\scriptsize{Tethering residues}}
\label{secttethering}

\scriptsize{The $C_\alpha$ atom of residue $46$ was
closest to the center of mass for WT SOD1 and all variants.
For mechanical scans,
residues 40, 45, 50, 54, and 55 were
either too close to the central $\Ca$, or were along the same beta
strand and gave anomalous force-extension profiles 
with artificially large forces probing covalent
bonding topology moreso than non-covalent stabilizing
interactions. In these cases,  the tethering residue
was moved to residue 76.
This always resolved the problem of large forces.
In assay (2) for dimer pulling simulations, the tethering point (on each of the
monomers) was $\Ca(46)$.

In assay (3) for Cu and Zn pulling simulations, the tethering point
was $\Ca(45)$ for Cu extraction and $\Ca(83)$ for Zn. The residue selection
was chosen by finding residues within $5$\AA~of the metal, and
selecting the residue that was the most buried (least solvent
accessible surface area (SASA)) as the tethering residue. The
resulting pathways are shown in the insets to
Figure~\ref{figCuZnFreedimer} (a). 
For the analysis of the (Cu,E) variant PDB 2R27, 
the Cu ion of chain A was shifted by 1.3 \AA~to a location associated with reduced
(Cu${}^{1+}$) that is closer to the putative Zn position. For this protein the
tethering residue was taken to be $\Ca(83)$.
}

\subsection{\scriptsize{Residues used for mechanical scans of protein stability}}
\label{secttethers}

\scriptsize{Every 5th residue along the protein sequence, including the first and
last residues, was taken as a tethering residue. As well, the central
positions of the predicted epitopes and anti-epitopes were added to
the data set~\cite{GCPalgorithm09}, as described in
reference~\cite{DasA12pnas1}.
If the center of the epitope happened to coincide with
a multiple of 5 already included in the scan (e.g. residue 10), then
the next residue (residue 11 in this case) was also taken. Thus 
a set of $48$ residues was used in the mechanical
scans: (1, 5, 10, 11,
15, 17, 20, 24, 25, 30, 31, 35, 38, 40, 45, 46, 50, 54, 55, 60,
65, 70, 73, 75, 80, 85, 90, 91, 95, 100, 101, 105, 108, 110, 115, 120,
121, 125, 130, 135, 136, 140, 141, 145, 146, 150, 151, 153). 
}

\subsection{\scriptsize{Proteins considered}}

\scriptsize{Proteins whose mechanical profiles were calculated are given in 
  Table~\ref{tabwham} below. For (Cu,E) SOD1, 2R27 was
used to generate an initial structure with the Cu shifted to the Zn
position (as it exists in the PDB structure), and a (Cu,E)
variant with the Cu in its putative position, by moving the Cu and
equilibrating the structure.

Several mutants were also considered for the calculation of metal
affinity and dimer affinity. These included
A4V, D124V, D125H, D76Y, D90A, G127X, G37R, G41D, G41S, G85R, G93A, G93C,
H43R, H46R, H46R/H48Q, H80R, I113T, L144F, L38V, S134N, T54R and
W32S. The corresponding PDB entries used to generate structures for
these mutants for simulations are given in Table~\ref{tabwham}.
}

\subsection{\scriptsize{Remodeling disordered/missing regions in PDB structures}}
\label{remodeling}

\scriptsize{(Cu,E) SOD1  (2R27) has two unstructured segments in the crystal
structure, residues 68-78 and 
residues 132-139, however a full length construct must be used
as an initial condition in simulations. The missing segments were added to the protein by the
following method: First, segments of protein corresponding to the missing
sequences were taken from the native holo SOD1 structure,  and Replica
Exchange Molecular Dynamics simulations of the free peptides in
explicit solvent~\cite{YaoY2008} were performed using the NAMD
simulation package~\cite{PhillipsJC05}, to obtain an ensemble of disordered conformations
for these two peptide fragments. Clustering was then performed on the
simulated ensemble for each peptide~\cite{SeeberM07}, and a conformation was chosen from
the largest cluster, such that its end-to-end distance best matched
the corresponding distance between the structured 
residues present in 2R27 that bracketed the missing sequence. 
After adding the missing loop segments, we ``back-mutated'' four
residues, A6C, S80H, S83D, and S111C, to recover the WT sequence,
using the PyMOL software package~\cite{DeLanoW2002}. 
The structure was then energy minimized, and equilibrated for $20$ ns, 
as described in section~\ref{sectsteeredmd}. 
A full
length construct was thus assembled to be used as an initial condition
in pulling simulations.  
}

\subsection{\scriptsize{Modeling proteins with no PDB structure}}
\label{modelingnopdb}

\scriptsize{Some proteins studied here had no PDB structural coordinates
available; see Table~\ref{tabwham}. For these proteins, a structure was built by modifying known PDB
structures of similar proteins. 
For example, disulfide-reduced WT SOD1 was created from (E,Zn)(SH) SOD1 (2AF2) by adding Cu to that structure,
and back-mutating A6C and S111C using the PyMOL software package~\cite{DeLanoW2002}.
Apo WT SOD1 was created from the apo SOD1 structure PDB 1RK7 by
back-mutating Q133E, E50F and E51G. 
Apo(SH) SOD1 was created from
the apo WT SOD1 structure 1RK7 by reducing the disulphide linkage
and mutating E50F, E51G, and Q133E.
The mutant H46R was generated from the H46R/H48Q double-mutant
of SOD1 by back-mutating Q48H. 
Similarly, the mutants W32S, G41S, G41D,
G93C, L144F, D90A and D76Y
were prepared from the holo SOD1 by mutating at the respective
positions.
The G127X mutant was created from holo SOD1 by deleting the
last 20 residues, and then mutating the last 6 residues of the
remaining sequence (KGGNEE) to correspond to the frame-shifted
non-native C-terminal peptide sequence (GGQRWK) prior to the termination
sequence.  
All modified
structures were energy minimized and equilibrated for 20 ns by
running an equilibrium simulation, before performing any pulling
assays.
}

\subsection{\scriptsize{Umbrella Sampling and Weighted Histogram Analysis Method (WHAM)}}

\subsubsection{\scriptsize{WHAM procedure for mechanical profiles}}
\label{whammethodmech}

\scriptsize{Initial configurations for use in WHAM were obtained from the pulling
simulations described in Methods section~\ref{sectsteeredmd}: $25$ initial
conditions between $0$ and $5$\AA~were
simulated for $10$ns each, in an umbrella potential with
stiffness $500$kJ/mol/nm$^2$ to constrain the simulations 
near their corresponding
separation distances. These $25$ simulations are then used to
reconstruct the free energy profile along the distance coordinate
using the weighted histogram analysis method (WHAM)~\cite{KumarS1992},
and the free energy difference between $0$ and $5$\AA~is then
obtained. A convergence check assured the result was unchanged if 30,
35 or 40 windows were used. 
}

\subsubsection{\scriptsize{WHAM procedure for metal expulsion and dimer separation}}
\label{whammethoddimermetal}

\scriptsize{To obtain free energies of metal binding, metals were
inserted into the putative binding locations for all SOD1 variants,
where they were found to be at least metastable. 
Dimer stabilities were calculated under the same metallation
conditions as the PDB structure for all mutants. Mutants with no
structure in the PDB were fully metallated. 

For the metal expulsion free energy calculations in Section~\ref{metalaffinity}, 
a most-direct straight-line path was first determined for the pulling
direction as described in Methods section~\ref{sectsteeredmd}, which
determined the tethering residue. 
The metal was pulled away from the protein using a spring constant of $500$
kJ/mol/nm$^2$ and a pull rate of $0.01$ nm/ps. 
For purposes of the free energy calculation, the final extension from
the equilibrium distance between metal and tethering residue
was taken to be approximately $30$\AA.
From these trajectories, snapshots were taken to generate the starting
configurations for the umbrella
sampling windows~\cite{PateyG1973,TorrieG1974,TorrieG1977}. 
An asymmetric distribution of
sampling windows was used, such that the window spacing was 
1\AA~between 0 and 20 \AA~separation, and 2\AA~beyond 20\AA~of 
separation. Such
spacing allowed for increasing detail at smaller separation distances, and
resulted in a total of 25 windows. In each window, 10 ns of MD was performed for
a total simulation time of 250 ns utilized for umbrella
sampling. Analysis of the results was performed using the weighted
histogram analysis method (WHAM)~\cite{KumarS1992}.  

To remove conformational distortion effects on the free energy, 
position restraints were applied in the following way. For a given 
$C_\alpha$ atom in the protein, all other $C_\alpha$ atoms within
$5$\AA~were constrained to have a roughly constant $C_\alpha$-$C_\alpha$
distance, the same as in the equilibrated structure, using spring
constants of $392\times 10^3$ kJ/mol/nm${}^2$.  The spring constant was varied to
interpolate between regimes where protein deformation is dominant, and
a regime where the constraining $C_\alpha$-$C_\alpha$ network can
potentially influence the expulsion free energy. 
Using $C_\alpha$ constraints allows the protein 
to retain its structure under force, 
while still allowing the side chains and partially the backbone to
fluctuate in response to the external perturbation. After metal
expulsion, $C_\alpha$ constraints were relaxed and
the relaxation free energy was calculated as described below. 

A similar procedure was employed for dimer separation. Here the
tethering points were taken to be the $C_\alpha$ atom of residue $46$
in each of the monomers. The monomers constituting the dimer were 
pulled away from each other, until the final distance was
approximately $30$\AA. The spring constant, pull rate, and window
separation were the same as in the metal expulsion analysis. Within
each monomer, $C_\alpha$ constraints were also applied to fix
the relative positions of the $C_\alpha$ atoms near their equilibrium
positions, and thus minimize conformational distortion during
monomerization. 

The free energy for metal expulsion or dimer separation was corrected
to account for protein relaxation in the final metal depleted or
monomerized state: $\Delta F_{tot} = \Delta F_{constr} + \Delta
F_{relax}$. That is, after either metal expulsion or  monomerization,
$C_\alpha$ constraints are gradually reduced from $392\times 10^3$
kJ/mol/nm${}^2$ to zero using 30 windows and 10ns of relaxation time
in each window, and free energy changes for this process are again
obtained using WHAM.  
Convergence of
the free energy values was tested by both varying the number of windows
(to 30, 35 and 45) and varying the length of the equilibrium
simulation in each window (to 15, 20, 25 and 30 ns). In all cases the
free energies were seen to have converged using the original
protocol.

To further confirm that equilibrium has been reached, thermodynamic
cycles were constructed for either re-insertion of the metal, or
re-dimerization of the monomers. The metal or dimer binding free
energies subject to $C_\alpha$ constraints were calculated, as well as
the subsequent relaxation free energies. Schematics of the thermodynamic cycles are
plotted in Figure~\ref{figthermo}, and values of each of the free energy
changes for WT protein and several mutants are given in
Table~S2 of the Supplementary Content. 
}

\subsubsection{\scriptsize{WHAM procedure for shifting Cu to the Zn-binding position}}
\label{whammethodshift}

\scriptsize{In the simulated conformations  of (Cu,E) SOD1 that have been equilibrated from the
crystal structure 2R27 (in which Cu is bound close to the putative Zn
position), Cu remains close to the Zn position and does not shift back
to the Cu binding site. The distance between the position 
of the Cu ion in the equilibrated structures from simulations of holo
WT SOD1 and the above (Cu,E) SOD1 is $7.92$ \AA. 
We calculated the free energy change to move Cu from its putative binding position 
to the Zn-position with the following method. We first pulled the Cu from its original position to 
the (Zn-binding) equilibrated position by applying a spring constant of $500$
kJ/mol/nm${}^2$ and a pulling speed of $10$ m/s, while enforcing
harmonic constraints on all of the $\Ca$ atoms as in Methods
Section~\ref{whammethoddimermetal}. 
Tethers were placed on the Cu ion and the $\Ca$ atom of residue $57$, 
because it is almost on the line joining the initial and final
positions of the Cu.
Conformations were collected after every $0.5$ \AA, and for each of the 
conformations a $10$ ns MD simulation was performed. These trajectories were 
used for umbrella sampling, and analysis of the results were performed using WHAM. 
}

\subsection{\scriptsize{Free energy differences from Jarzynski equalities, and
  finite sample size corrections}}
\label{freejarzynski}

\scriptsize{The free energy cost $\Delta F$ to extend a residue to $5$\AA~is found from
Jarzynski's equality
\begin{equation}
\Delta F =- kT \ln \left< \mbox{e}^{-W/kT} \right> \: ,
\label{eqj}
\end{equation}
where $W$ is the work from a non-equilibrium measurement, $kT$ is
Boltzmann's constant times the temperature in Kelvin, and the average
$\left< \cdots \right>$ is the ensemble average over replicated
pulling assays. 
For a finite sample size, the free energy in Eq.~(\ref{eqj}) becomes
\begin{equation}
\Delta F_{\mbox{\tiny J}} = - kT \ln \left< \mbox{e}^{-W/kT} \right>_N \\
= -kT \ln \left( \frac{1}{N} \sum_{i=1}^{N}
  \mbox{e}^{-W/kT} \right) \: ,
\label{eqjn}
\end{equation}
which must be corrected because of systematic bias for finite sample
size. Following Gore, Ritort, and Bustamante~\cite{GoreJ03}, an
estimate for the free energy change $\Delta F$ that accounts for
finite sample-size corrections is given by the following set of
equations: 
\begin{equation}
\begin{array}{l}
\displaystyle \Delta F = \Delta F_{\mbox{\tiny J}} -
\frac{\overline{W}_{dis2}}{N^{\alpha\left(\overline{W}_{dis2}\right)}} \\
\displaystyle \overline{W}_{dis2} = \overline{W}_{dis} + \frac{\overline{W}_{dis}
}{N^{\alpha\left(\overline{W}_{dis}\right)}}  \\
\displaystyle \overline{W}_{dis} = \left< W \right>_N - \Delta
F_{\mbox{\tiny J}} \\
\displaystyle \alpha\left(W\right) = \ln\left( 2 W/kT\right) / \ln\left[ C \left(
    \mbox{e}^{2W/kT} - 1\right)\right] \: .
\end{array} 
\label{eqjfinite}
\end{equation}
In the above equations, $\left< W \right>_N$ is the average work for
the $N$ pulling assays, and $\overline{W}_{dis}$ is the average
dissipated work- work that did not go into producing the free energy
change. The function $\alpha(W)$ is evaluated at two values of
dissipated work, $\overline{W}_{dis}$, and a corrected estimate for
the dissipated work, $\overline{W}_{dis2}$, which accounts for the
fact that $\overline{W}_{dis}$ is itself an underestimate because of the
finite sample size bias. The constant appearing in $\alpha(W)$
defines a crossover between small $N$ and large $N$ regimes and is not
rigorously determined. We take $C=15$ following Gore {\it et al}~\cite{GoreJ03}.

Table S6 in the Supplementary Content gives numerical values
for the various quantities in Equation~(\ref{eqjfinite}), in the assay where
residues $10$ and $17$ are pulled to 5\AA. The mean dissipated work 
$\overline{W}_{dis2}$ in our pulling assay is only about a kJ/mol,
indicating that pulling is near equilibrium, and finite sample-size corrections
are not large: about 1-3 kJ/mol. 
}

\subsection{\scriptsize{Statistical Analysis}}
\label{statanal}

\scriptsize{To compare whether the difference between two distinct cumulative
distributions is statistically signicant, the method described in
reference~\cite{DasA12pnas1} is used. A summary of the method
follows. 
The mechanical profile is a collection of $48$
work values for a given SOD1 variant, where each
work value has an error of $\sigma=2.7$kJ/mol. 
We wish to answer the following question:
given one work profile and its resultant cumulative distribution,
along with the error bars associated with 
the individual work values, what is the probability of obtaining a 
cumulative distributions at least as extreme as another given one by chance,
i.e. as arising from the original work profile?

To find the above probability and thus determine the
statistical significance of a given cumulative distribution, we employ
a ``Monte Carlo'' procedure of generating cumulative distributions from
a given ``baseline'' work profile. 
For example, to test the hypothesis that shifting the Cu from its
putative position to the Zn-binding position
mechanically stabilizes the (Cu,E) protein, we wish to find the probability
of obtaining a cumulative distribution at least as stabilized as
the (Cu-shift,E) cumulative distribution 
from the original (Cu,E) cumulative
distribution. The (Cu-shift,E) cumulative distribution has
$31$ work values that are more stable than those in the (Cu,E)
cumulative distribution; the difference in work values are plotted in Figure S2c,
largest to smallest. We seek cumulative distributions that have rank
ordered deviations at least as large as those of the
(Cu-shift,E) cumulative distribution. 
Work profiles are constructed by
adding gaussian noise of mean zero and standard deviation
$\sigma=2.7$~kJ/mol
to a the baseline work profile (here that of (Cu,E)
SOD1).
A cumulative distribution is then constructed for
each generated profile by sorting the values lowest to highest. After
constructing $N$ of such cumulative distributions, we ask whether one
has found a cumulative distribution with, e.g. rank-ordered positive
deviations at least as large as those in the (Cu-shift,E)
cumulative distribution. The value of N where the expected number of
trajectories is unity determines the statistical significance of the
difference between the two profiles:
$p=1/N$. 
Plots of the corresponding distributions for (Cu,E) SOD1 and
(Cu-shift,E) SOD1 are given in Figure S2 of the
Supplementary Content.}

\subsection{\scriptsize{Acknowledgements}}

\scriptsize{We thank Neil Cashman, Will Guest, 
Stephen Toope, Paul Whitford, and Michael Woodside for helpful and/or supportive
discussions. We also acknowledge funding from PrioNet Canada, and
acknowledge computational support from the WestGrid high-performance
computing consortium.} 

% --------------------------------------------------------------------

%% The Appendices part is started with the command \appendix;
%% appendix sections are then done as normal sections
%% \appendix

%% \section{}
%% \label{}

%% References
%%
%% Following citation commands can be used in the body text:
%% Usage of \cite is as follows:
%%   \cite{key}         ==>>  [#]
%%   \cite[chap. 2]{key} ==>> [#, chap. 2]
%%

%% References with bibTeX database:
\newpage
%\bibliographystyle{elsarticle-num}
%\bibliography{SSP}

%% Authors are advised to submit their bibtex database files. They are
%% requested to list a bibtex style file in the manuscript if they do
%% not want to use elsarticle-num.bst.

%% References without bibTeX database:

% \begin{thebibliography}{00}
%\begin{thebibliography}{10}
%   \input{arxiv-ms.bbl}
%\end{thebibliography}

%% \bibitem must have the following form:
%%   \bibitem{key}...
%%

% \bibitem{}

% \end{thebibliography}

\end{document}

\title{Mechanical probes of SOD1 predict systematic trends in metal
  and dimer affinity of ALS- associated mutants (Supplementary Content)}

\author{Atanu Das\affil{1}{Department of Physics and Astronomy, University of British Columbia, Vancouver,
Canada}
\and
Steven S. Plotkin\affil{1}{}}

\maketitle
\thispagestyle{empty}
\pagestyle{empty}
{\Large {\bf Supplemental Figures}} \\

\renewcommand{\thefigure}{S\arabic{figure}}
\begin{figure}[H]
\centering
\includegraphics*[width=16cm]{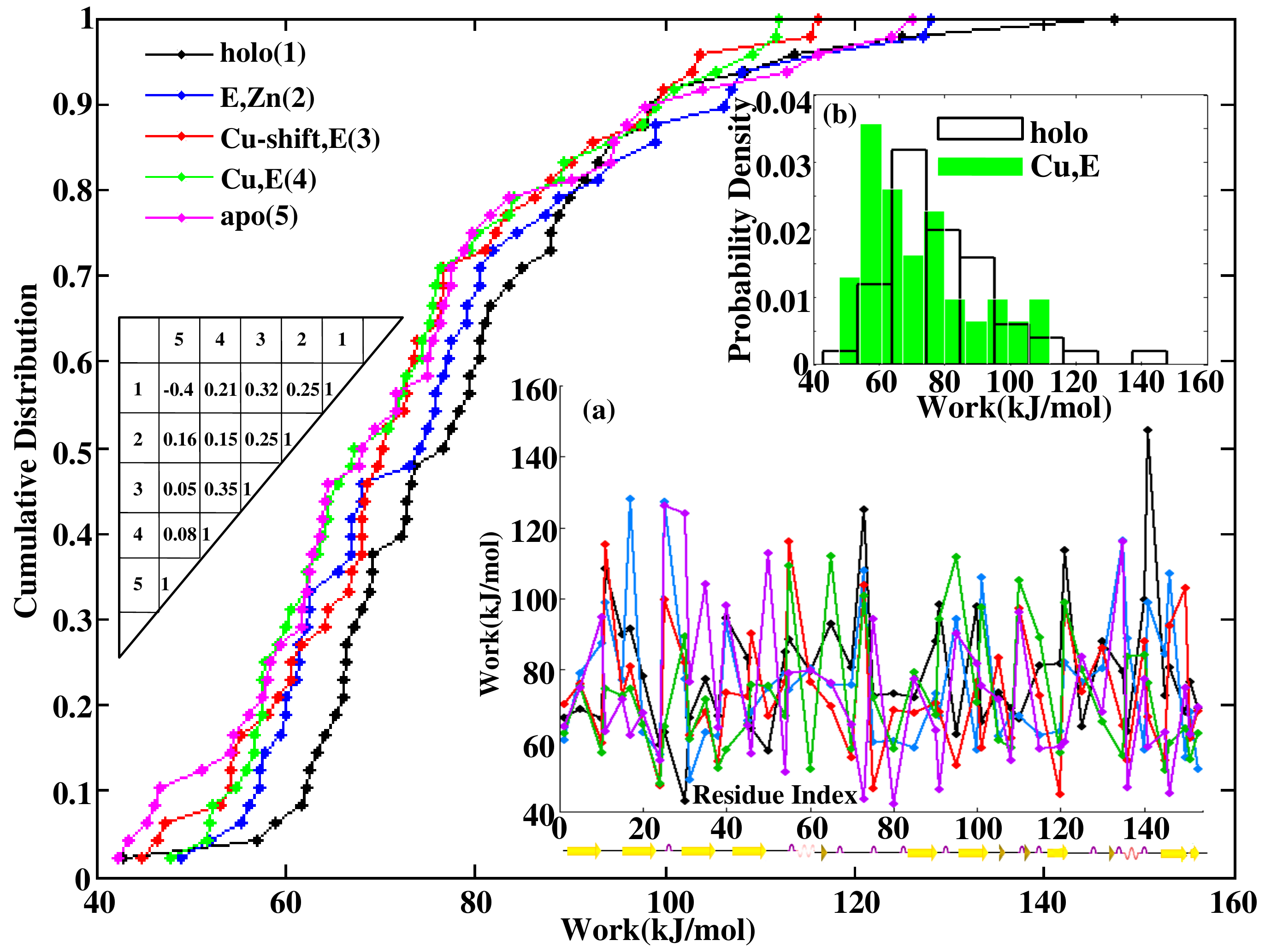}
\caption[]{
{\small The effect of metal depletion on the mechanical stability of
SOD1. (Inset panel a) Mechanical profiles of holo SOD1, (E,Zn)
SOD1, (Cu,E) SOD1, and Zn-depleted SOD1 with the Cu ion shifted
from its putative position to that of the Zn (Cu-shift,E), which is
near the location of
the Cu observed in the crystal structure of PDB
% 2R27~\cite{RobertsBR07}. Profiles are
2R27. Profiles are
color-coded according to the legend in the upper left. The effect of
metal depletion modulates the entire mechanical
stability profile.
Work values for individual residues are given in Table S5 of the
Supplementary Content.
(Inset b) Probability distributions of mechanical stability for holo
and (Cu,E) variants. This representation shows that the
distributions are indeed different, however there is significant
overlap, which along with the necessity of binning the data to
construct the histogram, makes the differences difficult to
quantify. (Main Panel) Cumulative histograms of the above SOD1
variants. The cumulative histogram requires no binning, and simply
rank orders and accumulates the work values. Using this
representation, discrepancies between work profiles most clearly
emerge. The mechanical stability of the SOD1 variants can be rank
ordered, strongest to weakest, as holo, (E,Zn),
(Cu-shift,E), and (Cu,E). The distribution of apo SOD1
is broadened compared to (Cu,E) SOD1, but not overall
weakened.
Also shown in the main panel is a correlation matrix resulting from a
pairwise comparison of the work values for all pairings of SOD1
variants. The  profiles do not correlate, again supporting the notion
that metal depletion modulates the entire mechanical stability
profile. }}
\label{figsstest}
\end{figure}

\renewcommand{\thefigure}{S\arabic{figure}}
\begin{figure}[H]
\centering
\includegraphics*[width=16cm]{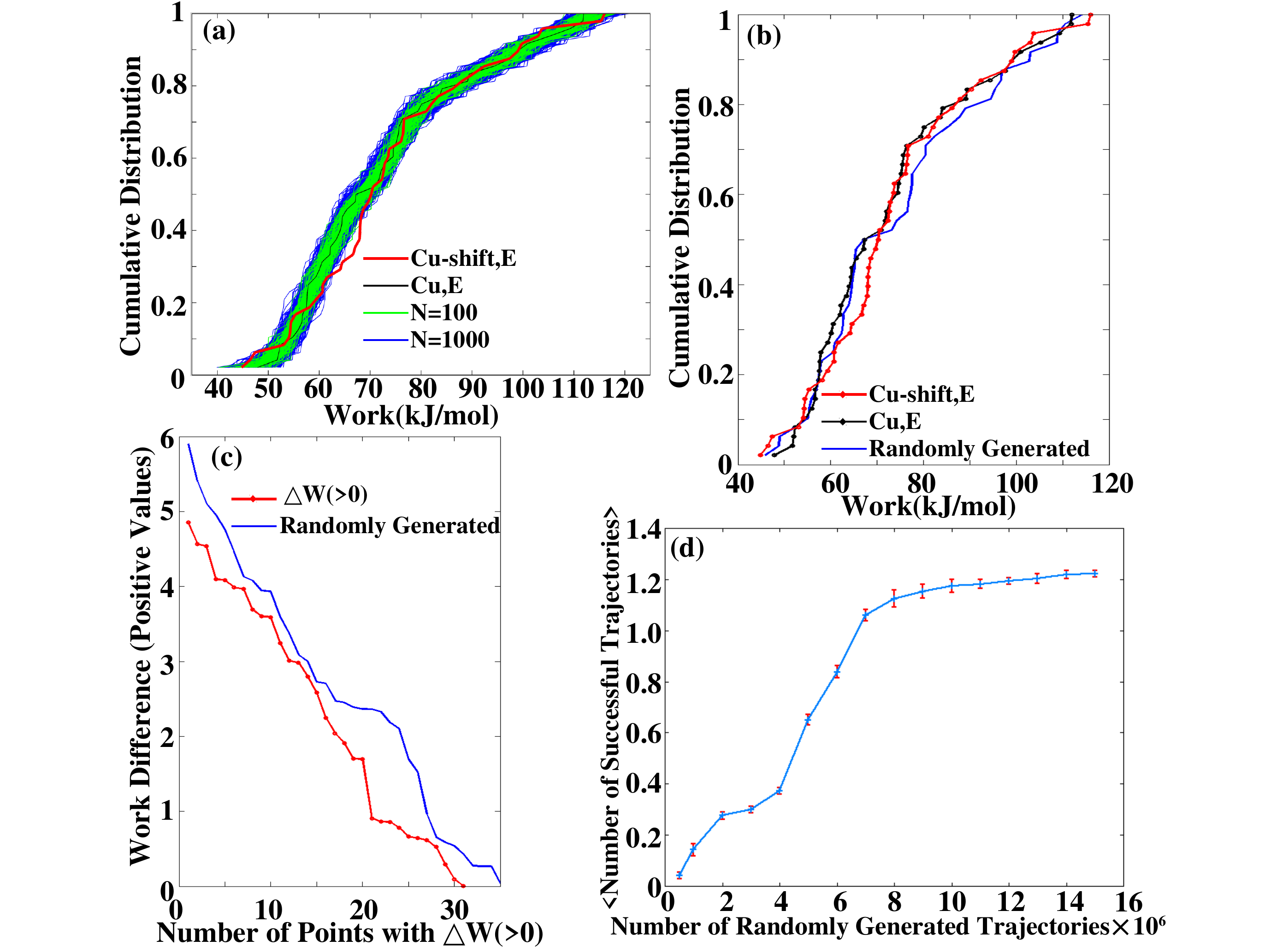}
\caption[]{
{\small (a) Statistical significance test between (Cu,E) and (Cu-shift,E)
cumulative distributions. The black curve denotes the (Cu,E)
profile, and the red curve denotes the (Cu-shift,E)
profile. The collection of green lines corresponds to 100
randomly-generated cumulative distributions in the following way:
starting from the (Cu,E) work values, gaussian noise with zero
mean and standard deviation $\sigma = 2.7$~kJ/mol  is added to
each work value. The resulting collection of work values are then
rank-ordered to generate a cumulative distribution, and this process
is then repeated $N$ times. The collection of blue lines
corresponds to $N=1000$. For large enough $N$, one will begin to find
outlying cumulative distributions which deviate as much or more than a
given trial distribution, here (Cu-shift,E). An example of
such an outlier is shown in panel (b). 
In this case, the number of positive deviations (differences in work
value) between the red and
black curves in panel (a) are recorded (31), along with their work
values, which are themselves rank ordered largest to smallest and
plotted in figure (c). An outlier at least as extreme as
(Cu-shift,E) must have a set of positive work deviations
such that the corresponding  curve lies above the red curve in panel
(c). In this case the randomly generated outlier has 36 positive
deviations in work, 31 of which are larger than those between
(Cu-shift,E) and (Cu,E) variants. 
The number of randomly generated trajectories is increased, until the
mean number of successful trajectories, corresponding to the criterion
in panel (c), exceeds unity. A plot of the mean number of successful
trajectories as a function of $N$ is given in panel (d). The
statistical significance $p$ of a given trajectory is then given as
$p=1/N_1$, where $N_1$ is the number of randomly generated
trajectories that give an expectation value of unity for the number of
successfully-generated outliers. 
In this case, the statistical significance of (Cu-shift,E)
is about $1/(7\times 10^6)$, or about $1.4\times 10^{-7}$. }
}
\label{figsstest}
\end{figure}

\renewcommand{\thefigure}{S\arabic{figure}}
\begin{figure}[H]
\centering
\includegraphics*[width=16cm]{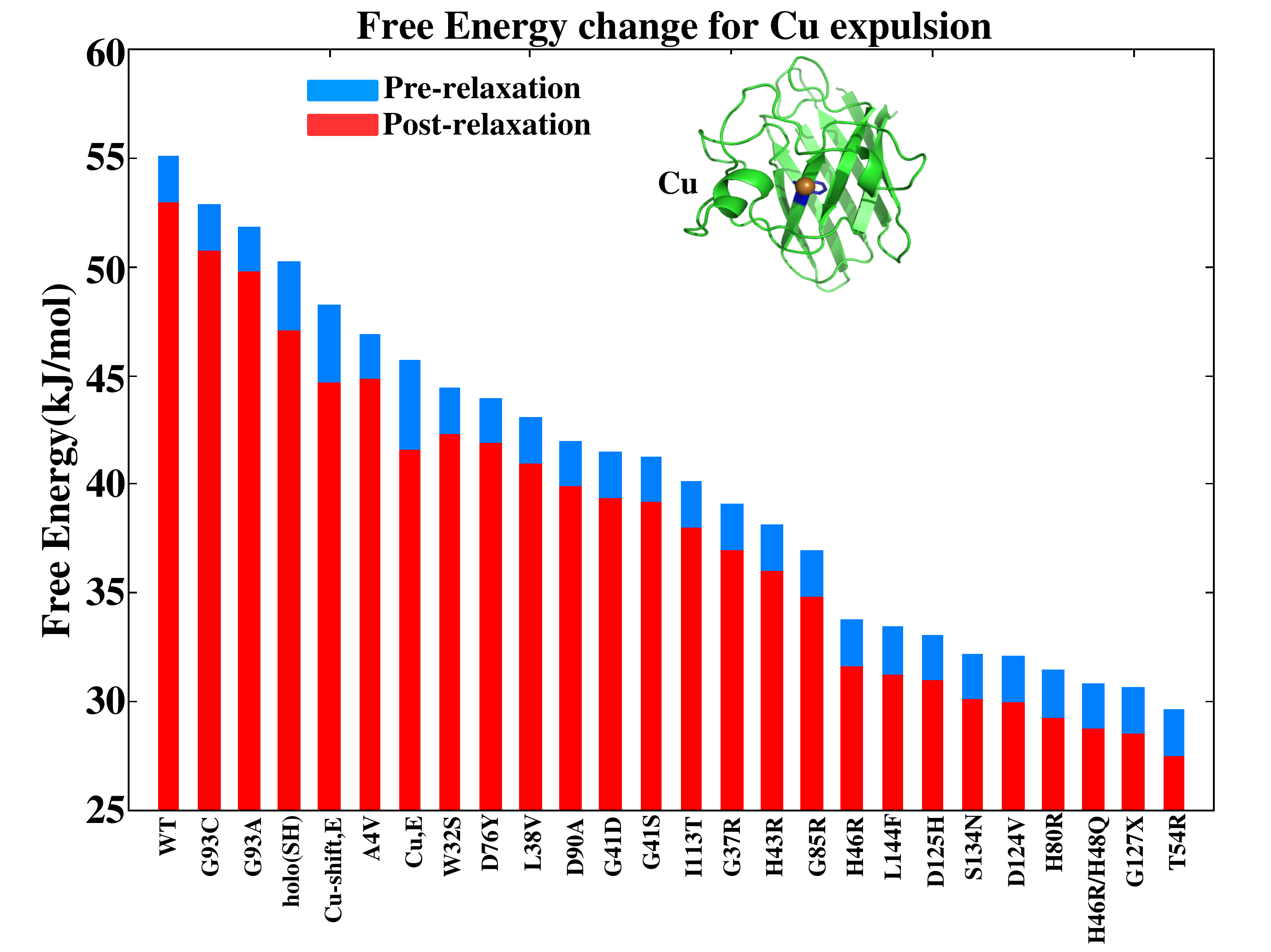}
\caption[]{ {\small Free energy change for Cu expulsion, for WT SOD1 along with
  several mutants and post-translationally modified variants. The
  heights of the blue
  bars indicate the corresponding free energies of metal binding before
  conformational relaxation of the protein in the unbound state. Red
  bar heights indicate
  the corresponding free energies of binding after relaxation of the
  protein in the unbound state.  A decrease in free energy value is observed for all
the proteins.}
}
\label{figrelaxcu}
\end{figure}

\renewcommand{\thefigure}{S\arabic{figure}}
\begin{figure}[H]
\centering
\includegraphics*[width=16cm]{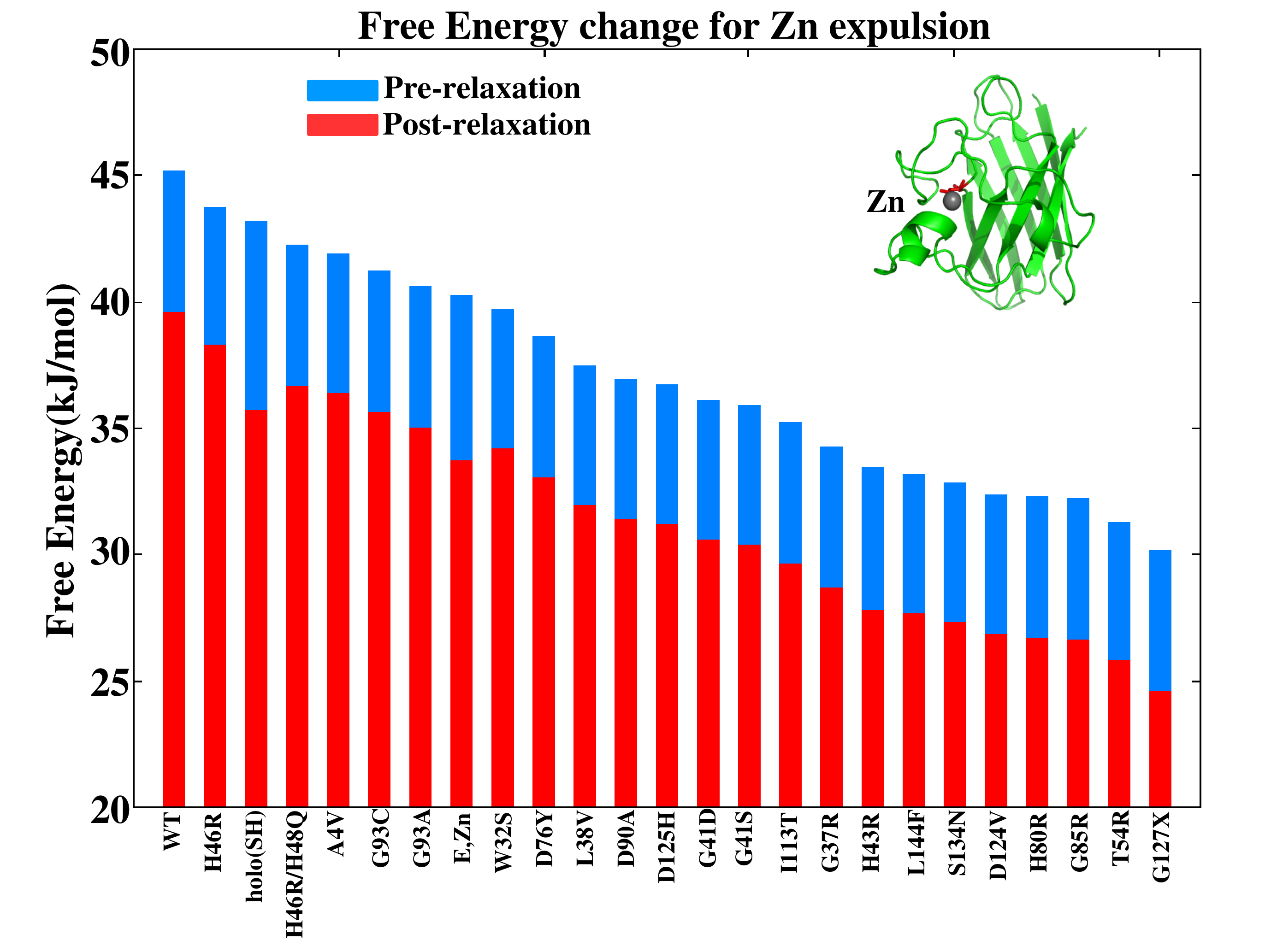}
\caption[]{{\small Free energy change for Zn expulsion, for WT SOD1 along with
  several mutants and post-translationally modified variants. The
  heights of the blue
  bars indicate the corresponding free energies of metal binding before
  conformational relaxation of the protein in the unbound state. Red
  bar heights indicate
  the corresponding free energies of binding after relaxation of the
  protein in the unbound state.  }
}
\label{figrelaxzn}
\end{figure}

\renewcommand{\thefigure}{S\arabic{figure}}
\begin{figure}[H]
\centering
\includegraphics*[width=16cm]{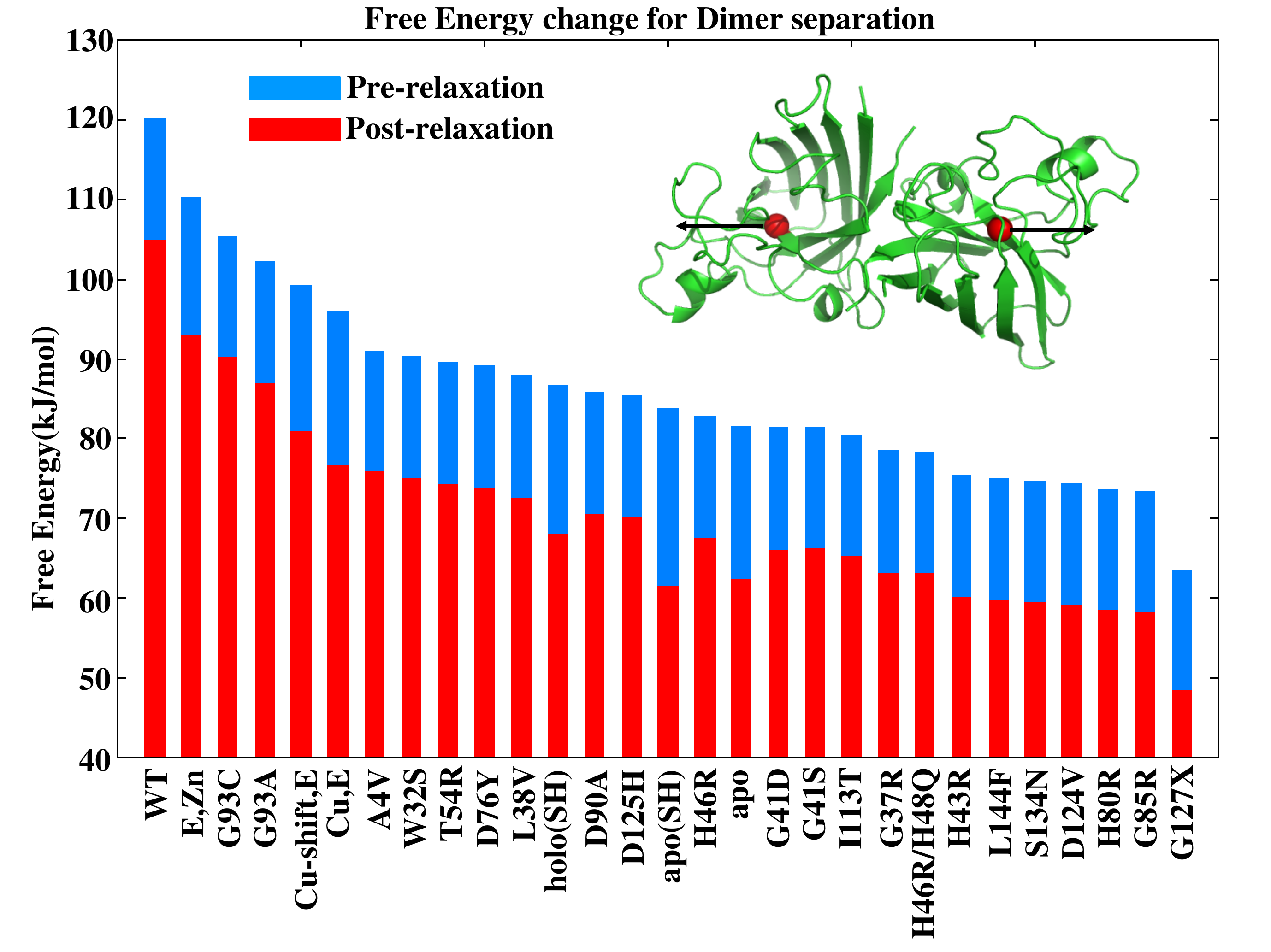}
\caption[]{
{\small Free energy change for dimer separation, for WT holo SOD1 along with
  several mutants and post-translationally modified variants. All
  mutants plotted here are taken in the holo state. The
  heights of the blue
  bars indicate the corresponding free energy to monomerize the dimer before
  conformational relaxation of both proteins in the unbound state. Red
  bar heights indicate
  the corresponding free energies of binding after conformational relaxation of both
  protein in the monomerized state.}
}
\label{figrelaxdimer}
\end{figure}

\renewcommand{\thefigure}{S\arabic{figure}}
\begin{figure}[H]
\centering
\includegraphics*[width=16cm]{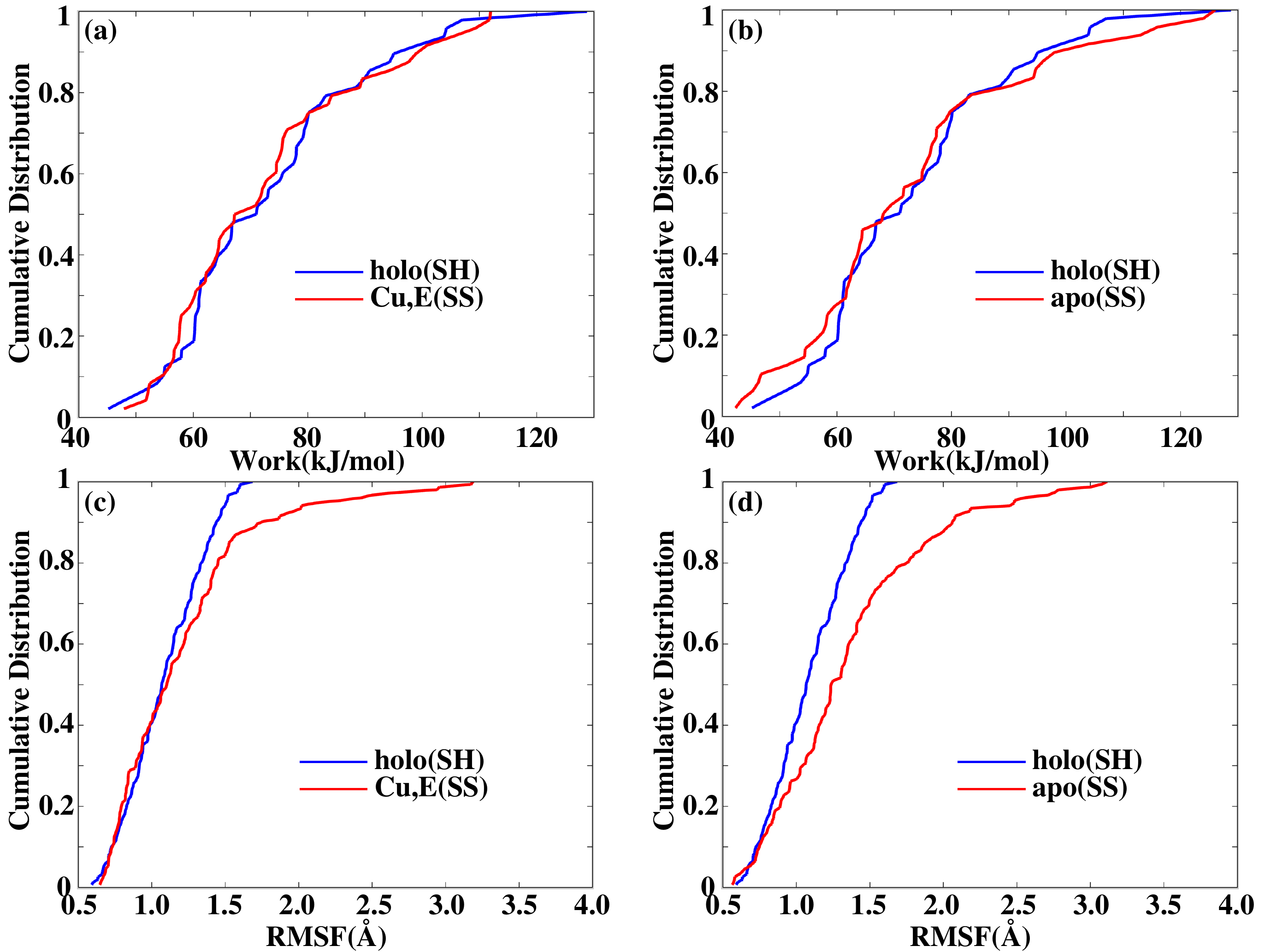}
\caption[]{
{\small Loss of Zn is at least as destabilizing to SOD1 mechanical
  stability as disulfide reduction. Panel (a) gives the cumulative distribution
of the work profiles for holo(SH) and Cu,E(SS) SOD1 variants, which
shows that Cu,E(SS) SOD1 has marginally weaker mechanical stability
than holo(SH) SOD1 ($p=1.4$e-$7$). This trend
is also observed in the cumulative distribution of the equilibrium
fluctuations of the holo(SH) and Cu,E(SS) variants (Panel (c)), where
Cu,E(SS) shows marginally larger fluctuations.  
Panel (b) compares cumulative distributions of work profiles for
holo(SH) and apo(SS) SOD1, showing apo(SS) SOD1 is marginally weaker
mechanically ($p=3$e-$8$). The cumulative distribution of equilibrium
fluctuations of these two variants (panel d) also shows that apo(SS) SOD1 has
more significant dynamics in the native state. 
% distributions are plotted in panel (c). Depletion of both the metals
% while retaining the disulfide bond has moderately
% higher destabilizing effect on SOD1 mechanical stability than disulfide
% reduction of holo SOD1 as evident from cumulative distributions of work values of apo(SS) and
% holo(SH) variants(panel(b)). This is further validated by the cumulative distributions
% of the RMSF values of those variants(panel(d)).
} }
\label{figpred}
\end{figure}

\renewcommand{\thefigure}{S\arabic{figure}}
\begin{figure}[H]
\centering
\includegraphics*[width=16cm]{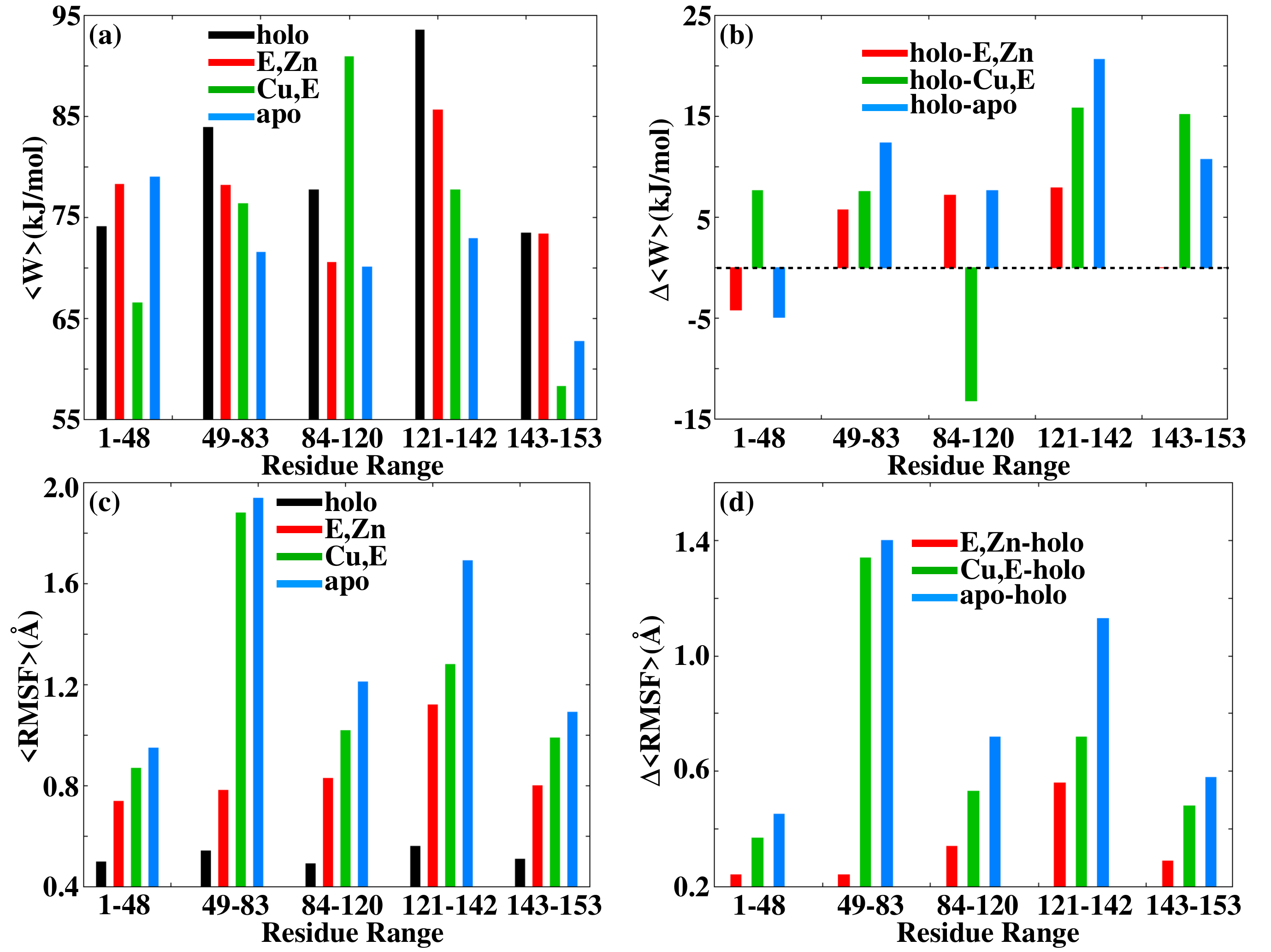}
\caption[]{
{\small Zn-binding and electrostatic loops (ZBL and ESL) show decreased mechanical
  rigidity and increased dynamic mobility upon metal release. The
SOD1 sequence is divided into 5 different regions:
the region N-terminal to the ZBL (residues 1-48), the ZBL (residues 49-83),
the middle region of sequence between the ZBL and ESL (residues
84-120), the ESL (residues 121-142), and the C-terminal
region after the ESL (residues 143-153). 
Panel (a) gives the average work values (as calculated by pulling to
5\AA) for those regions, for holo, (E,Zn), (Cu,E) and apo SOD1
variants. 
Panel (b) gives the difference in average work
values of those 5 regions in the various metal-deficient forms of SOD1
from holo SOD1 (e.g. $\left< W \right>_{holo} - \left< W
\right>_{apo}$). 
This shows that metal loss systematically destabilizes the ZBL and
ESL; other regions of the proteins show less systematic trends. 
Panel (c) gives the root mean squared fluctuations (RMSF) averaged
over the above 5 regions, and panel (d) gives the difference in RMSF
of those 5 regions in the various metal-deficient forms of SOD1
from holo SOD1 (e.g. $\left< RMSF \right>_{apo} - \left< RMSF
\right>_{holo}$). 
This recapitulates the mechanical stability assay: metal loss
systematically increases dynamic fluctuations in the ZBL and
ESL. Other regions show weaker trends in dynamic fluctuations. 
} }
\label{figmetalloss}
\end{figure}

\newpage

{\Large {\bf Supplemental Tables}}

\renewcommand{\thetable}{S\arabic{table}}
\renewcommand{\arraystretch}{1}
\begin{table}[H]
\caption{Free energy of Cu, Zn expulsion and Dimer separation (kJ/mol)}
\begin{tabular}{ccccc}
\hline
SOD1 variants & Cu${}^2$${}^+$ & Cu${}^1$${}^+$ & Zn${}^2$${}^+$ & Dimer \\
\hline
WT & $52.9$ & $26.4$ & $39.6$ & $104.9$ \\
holo(SH) & $47.1$ & $23.5$ & $35.7$ & $68.1$ \\
(Cu-shift,E) & $44.6$ & $22.3$ & - & $80.9$ \\
(Cu,E) & $41.5$ & $20.7$ & - & $76.6$ \\
(E,Zn) & - & - & $33.7$ & $93.1$ \\
(E,Zn)(SH) & - & - & $32.6$ & $66.7$ \\
apo(SH) & - & - & - & $61.5$ \\
apo & - & - & - & $62.3$ \\
G$93$C & $50.7$ & $25.3$ & $35.6$ & $90.2$ \\
G$93$A & $49.7$ & $24.8$ & $35.0$ & $86.8$ \\
A$4$V & $44.8$ & $22.4$ & $36.4$ & $75.7$ \\
W$32$S & $42.3$ & $21.2$ & $34.1$ & $74.9$ \\
D$76$Y & $41.8$ & $20.9$ & $33.0$ & $73.8$ \\
L$38$V & $40.9$ & $20.4$ & $31.9$ & $72.6$ \\
D$90$A & $39.9$ & $19.9$ & $31.4$ & $70.5$ \\
G$41$D & $39.3$ & $19.6$ & $30.5$ & $66.0$ \\
G$41$S & $39.1$ & $19.5$ & $30.4$ & $66.1$ \\
I$113$T & $38.0$ & $19.0$ & $29.6$ & $65.1$ \\
G$37$R & $36.9$ & $18.4$ & $28.7$ & $63.2$ \\
H$43$R & $35.9$ & $17.9$ & $27.8$ & $60.1$ \\
G$85$R & $34.8$ & $17.3$ & $26.6$ & $58.2$ \\
H$46$R & $31.6$ & $15.8$ & $38.2$ & $67.4$ \\
L$144$F & $31.2$ & $15.6$ & $27.6$ & $59.6$ \\
D$125$H & $30.9$ & $15.4$ & $31.2$ & $70.1$ \\
S$134$N & $30.1$ & $15.0$ & $27.3$ & $59.5$ \\
D$124$V & $30.0$ & $14.9$ & $26.8$ & $59.0$ \\
H$80$R & $29.2$ & $14.6$ & $26.7$ & $58.3$ \\
H$46$R/H$48$Q & $28.7$ & $14.3$ & $36.6$ & $63.0$ \\
G$127$X & $28.5$ & $14.2$ & $24.6$ & $48.3$ \\
T$54$R & $27.4$ & $13.7$ & $25.8$ & $74.1$ \\
(E,Zn)H$46$R & - & - & - & $60.2$ \\
(E,Zn)H$46$R/H$48$Q & - & - & - & $58.9$ \\
apoA$4$V & - & - & - & $47.1$ \\
apoG$37$R & - & - & - & $51.6$ \\
apoL$38$V & - & - & - & $48.1$ \\
apoG$41$S & - & - & - & $47.1$ \\
apoH$46$R & - & - & - & $49.1$ \\
apoT$54$R & - & - & - & $48.5$ \\
apoD$76$Y & - & - & - & $49.1$ \\
apoG$85$R & - & - & - & $47.9$ \\
apoD$90$A & - & - & - & $50.1$ \\
apoG$93$C & - & - & - & $50.9$ \\
apoI$113$T & - & - & - & $46.9$ \\
apoD$124$V & - & - & - & $47.0$ \\
apoD$125$H & - & - & - & $48.0$ \\
apoG$127$X & - & - & - & $44.2$ \\
apoS$134$N & - & - & - & $48.1$ \\
apoL$144$F & - & - & - & $49.1$ \\
\hline
\label{tabfree}
\end{tabular}
\end{table}

\renewcommand{\thetable}{S\arabic{table}}
\renewcommand{\arraystretch}{1}
\begin{table}[H]
\caption{Free energy change for metal expulsion and dimer separation (kJ/mol) -
  thermodynamic cycle (see Figure 4, main text)}
\begin{tabular}{ccccc}
\hline
Metal & $\Delta F_1^{constr}$ & $-\Delta F_2^{relax}$ & $\Delta F_3^{constr}$ & $-\Delta F_4^{relax}$ \\
\hline
WT(Cu) & $55.0$ & $2.1$ & $53.5$ & $0.8$ \\
WT(Zn) & $45.1$ & $5.5$ & $41.4$ & $1.9$ \\
Cu,E(Cu) & $45.6$ & $4.1$ & $43.4$ & $1.7$ \\
E,Zn(Zn) & $40.2$ & $6.5$ & $36.6$ & $3.0$ \\
A4V(Cu) & $46.9$ & $2.1$ & $45.5$ & $0.9$ \\
A4V(Zn) & $41.9$ & $5.5$ & $38.6$ & $2.1$ \\
G93A(Cu) & $51.8$ & $2.1$ & $50.5$ & $0.9$ \\
G93A(Zn) & $40.6$ & $5.6$ & $37.4$ & $2.2$ \\
\hline
Dimer & $\Delta F_1^{constr}$ & $-\Delta F_2^{relax}$ & $\Delta F_3^{constr}$ & $-\Delta F_4^{relax}$ \\
\hline
WT & $120.1$ & $15.2$ & $111.7$ & $6.6$ \\
Cu,E & $95.8$ & $19.2$ & $86.0$ & $9.2$ \\
E,Zn & $110.2$ & $17.1$ & $100.2$ & $7.2$ \\
A4V & $91.0$ & $15.3$ & $84.9$ & $8.9$ \\
G93A & $102.1$ & $15.3$ & $93.7$ & $7.1$ \\
\hline
\label{tabthermocycle}
\end{tabular}
\end{table}

\renewcommand{\thetable}{S\arabic{table}}
\renewcommand{\arraystretch}{1}
\begin{table}[H]
\label{tabworkrmsf}
\caption{Correlation between Work and RMSF of SOD1 variants}
\begin{tabular}{cccccccc}
\hline
Correlation & holo & E,Zn & Cu-shift,E & Cu,E & apo & holo(SH) & apo(SH) \\
\hline
r & $0.06$ & $-0.02$ & $0.06$ & $0.08$ & $0.01$ & $0.07$ & $4.7\times10^{-5}$ \\
p & $0.69$ & $0.89$ & $0.66$ & $0.57$ & $0.94$ & $0.61$ & $0.99$ \\
\hline
\end{tabular}
\end{table}

\renewcommand{\thetable}{S\arabic{table}}
\renewcommand{\arraystretch}{1}
\begin{table}[H]
\tiny
\caption{Correlation table of work values for different variants
of SOD1}
\begin{tabular}{c|c|c|c|c|c|c|c}
\hline
\backslashbox{p} {r${}^\dag$} & holo & E,Zn & Cu-shift,E & Cu,E & apo & holo(SH) & apo(SH) \\
\hline
holo & \backslashbox{1} {1} & 0.25 & 0.32 & 0.21 & -0.37 & -0.02 & -0.16 \\
\hline
E,Zn & 0.09 & \backslashbox{1} {1} & 0.25 & 0.15 & 0.16 & -0.06 & -0.07 \\
\hline
Cu-shift,E & 0.03 & 0.09 & \backslashbox{1} {1} & 0.35 & 0.05 & 0.17 & 0.05 \\
\hline
Cu,E & 0.15 & 0.30 & 0.01 & \backslashbox{1} {1} & 0.08 & 0.13 & 0.07 \\
\hline
apo & 0.01 & 0.26 & 0.74 & 0.57 & \backslashbox{1} {1} & 0.30 & 0.51 \\
\hline
holo(SH) & 0.91 & 0.68 & 0.26 & 0.38 & 0.04 & \backslashbox{1} {1} & 0.33 \\
\hline
apo(SH) & 0.27 & 0.63 & 0.72 & 0.62 & 0.0002 & 0.02 & \backslashbox{1} {1} \\
\hline
\label{tabcorr}
\end{tabular}
\footnotesize{${}^\dag \! r$ indicates the correlation coefficient between
  work profiles of different SOD1 variants, and $p$ indicates the
  statistical significance of the correlation coefficient.} 
\end{table}

\renewcommand{\thetable}{S\arabic{table}}
\renewcommand{\arraystretch}{1}
\begin{table}[H]
\label{tabE}
\caption{Mechanical work values of WT SOD1 variants - the values reported in the table
depict the work in kJ/mol needed to pull a particular residue up to 5 \AA. }
\begin{tabular}{cccccccc}
\hline
Res.Ind. & holo & (E,Zn) & (Cu-shift,E) & (Cu,E) & apo & holo(SH) & apo(SH) \\
\hline
$1$ & $66.3$ & $60.1$ & $70.2$ & $62.2$ & $63.9$ & $60.1$ & $60.3$ \\
$5$ & $68.8$ & $79.0$ & $76.0$ & $75.1$ & $74.9$ & $77.5$ & $88.9$ \\
$10$ & $66.1$ & $87.4$ & $59.3$ & $56.7$ & $94.6$ & $79.4$ & $81.2$ \\
$11$ & $108.4$ & $98.9$ & $115.3$ & $74.5$ & $62.5$ & $63.8$ & $59.5$ \\
$15$ & $89.9$ & $77.1$ & $72.8$ & $72.8$ & $71.5$ & $79.1$ & $68.8$ \\ \hline
$17$ & $91.5$ & $128.0$ & $81.0$ & $74.5$ & $61.6$ & $61.2$ & $45.1$ \\
$20$ & $78.2$ & $62.4$ & $64.5$ & $64.5$ & $67.7$ & $103.8$ & $60.6$ \\
$24$ & $58.8$ & $56.0$ & $47.4$ & $47.8$ & $54.4$ & $48.0$ & $56.7$ \\
$25$ & $62.3$ & $127.3$ & $99.8$ & $64.0$ & $126.0$ & $62.9$ & $87.3$ \\
$30$ & $42.9$ & $77.3$ & $82.0$ & $89.4$ & $124.0$ & $75.7$ & $113.4$ \\ \hline
$31$ & $66.3$ & $48.9$ & $61.5$ & $60.5$ & $76.5$ & $60.2$ & $55.9$ \\
$35$ & $77.3$ & $62.2$ & $68.0$ & $71.7$ & $104.0$ & $66.7$ & $84.1$ \\
$38$ & $67.1$ & $61.2$ & $54.1$ & $52.4$ & $63.5$ & $64.2$ & $64.7$ \\
$40$ & $94.4$ & $92.9$ & $73.5$ & $57.4$ & $98.0$ & $83.1$ & $91.5$ \\
$45$ & $83.4$ & $65.5$ & $72.3$ & $64.4$ & $64.3$ & $78.0$ & $68.1$ \\ \hline
$46$ & $63.3$ & $68.0$ & $90.2$ & $75.7$ & $56.2$ & $71.0$ & $66.8$ \\
$50$ & $57.0$ & $74.9$ & $67.0$ & $75.5$ & $112.8$ & $78.0$ & $78.2$ \\
$54$ & $85.0$ & $78.9$ & $76.4$ & $67.0$ & $51.1$ & $45.2$ & $71.6$ \\
$55$ & $88.6$ & $74.2$ & $116.0$ & $109.2$ & $78.9$ & $94.3$ & $72.6$ \\
$60$ & $79.5$ & $80.4$ & $76.5$ & $52.1$ & $79.7$ & $90.8$ & $58.4$ \\ \hline
$65$ & $92.9$ & $75.7$ & $69.6$ & $111.9$ & $76.2$ & $107.0$ & $83.2$ \\
$70$ & $80.5$ & $75.7$ & $55.2$ & $57.6$ & $64.4$ & $57.8$ & $95.8$ \\
$73$ & $125.0$ & $108.0$ & $103.7$ & $100.8$ & $43.4$ & $55.0$ & $60.8$ \\
$75$ & $72.8$ & $59.5$ & $46.4$ & $72.1$ & $94.3$ & $53.6$ & $71.4$ \\
$80$ & $73.1$ & $60.0$ & $68.6$ & $57.6$ & $42.3$ & $73.0$ & $72.1$ \\ \hline
$85$ & $72.1$ & $57.9$ & $67.9$ & $79.3$ & $77.3$ & $65.6$ & $70.8$ \\
$90$ & $87.9$ & $73.1$ & $70.5$ & $67.2$ & $62.9$ & $128.8$ & $74.4$ \\
$91$ & $98.4$ & $66.9$ & $68.0$ & $94.3$ & $46.2$ & $60.2$ & $61.3$ \\
$95$ & $61.7$ & $94.2$ & $53.0$ & $111.7$ & $90.1$ & $89.8$ & $82.9$ \\
$100$ & $97.8$ & $57.3$ & $76.6$ & $70.8$ & $81.6$ & $88.4$ & $87.4$ \\ \hline
$101$ & $65.3$ & $106.0$ & $58.1$ & $97.6$ & $75.5$ & $61.2$ & $40.2$ \\
$105$ & $73.5$ & $61.3$ & $83.3$ & $60.0$ & $71.6$ & $95.0$ & $63.5$ \\
$108$ & $69.2$ & $67.0$ & $60.6$ & $57.8$ & $54.3$ & $51.0$ & $40.2$ \\
$110$ & $66.2$ & $67.0$ & $97.3$ & $105.2$ & $96.0$ & $73.2$ & $86.6$ \\
$115$ & $81.1$ & $61.5$ & $72.6$ & $89.1$ & $57.6$ & $82.2$ & $72.7$ \\ \hline
$120$ & $81.6$ & $62.6$ & $44.7$ & $56.6$ & $58.1$ & $66.8$ & $101.1$ \\
$121$ & $113.6$ & $81.9$ & $99.0$ & $98.9$ & $59.5$ & $66.5$ & $66.7$ \\
$125$ & $64.1$ & $76.6$ & $73.7$ & $80.1$ & $83.5$ & $75.1$ & $82.8$ \\
$130$ & $87.9$ & $80.4$ & $86.1$ & $65.4$ & $68.1$ & $80.1$ & $103.1$ \\
$135$ & $79.4$ & $116.2$ & $64.1$ & $55.8$ & $116.0$ & $104.2$ & $85.1$ \\ \hline
$136$ & $62.5$ & $88.8$ & $54.4$ & $83.6$ & $46.7$ & $61.0$ & $70.2$ \\
$140$ & $99.7$ & $57.3$ & $87.9$ & $84.1$ & $77.4$ & $79.9$ & $92.1$ \\
$141$ & $147.3$ & $98.9$ & $66.6$ & $76.3$ & $58.3$ & $54.8$ & $56.8$ \\
$145$ & $72.8$ & $84.4$ & $54.3$ & $51.7$ & $62.2$ & $71.2$ & $76.2$ \\
$146$ & $80.5$ & $107.0$ & $92.4$ & $59.3$ & $45.3$ & $61.0$ & $61.2$ \\ \hline
$150$ & $68.1$ & $55.3$ & $103.0$ & $63.3$ & $74.8$ & $99.2$ & $80.2$ \\
$151$ & $76.6$ & $68.0$ & $60.6$ & $54.7$ & $61.7$ & $58.0$ & $66.1$ \\
$153$ & $69.0$ & $52.1$ & $68.2$ & $62.1$ & $69.5$ & $60.4$ & $60.3$ \\
\hline
\end{tabular}
\end{table}

\renewcommand{\thetable}{S\arabic{table}}
\renewcommand{\arraystretch}{1}
\begin{table}[H]
\label{tabfinite}
\caption{Finite sample size correction of Jarzynski equality (energies
are in kJ/mol)}
\begin{tabular}{cccccccc}
\hline
Residue Index & $\left< W \right>_N$ & $\overline{W}_{dis}$ & $\overline{W}_{dis2}$ & $\alpha(\overline{W}_{dis})$ & $\alpha(\overline{W}_{dis2})$ & $ \Delta F_{\mbox{\tiny J}} $ & $\Delta F $ \\
\hline
$10$ & $56.24$ & $0.110$ & $1.33$ & $-0.804$ & $0.25$ & $56.13$ & $55.50$ \\
$17$ & $80.27$ & $0.108$ & $1.13$ & $-0.756$ & $0.28$ & $80.16$ & $77.55$ \\
\hline
\end{tabular}
\end{table}

% \newpage
% \bibliographystyle{elsarticle-num}
% \bibliography{SSP}